\documentclass[sn-mathphys,Numbered]{sn-jnl}
\usepackage{array}
\usepackage{graphicx}
\usepackage{float}
\usepackage{multirow}%
\usepackage{amsmath,amssymb,amsfonts}
\usepackage{amsthm}
\usepackage{mathrsfs}
\usepackage[title]{appendix}
\usepackage{xcolor}
\usepackage{textcomp}
\usepackage{manyfoot}%
\usepackage{booktabs}%
\usepackage{algorithm}
\usepackage{mathtools}
\usepackage{algorithmicx}%
\usepackage{algpseudocode}%
\usepackage{listings}%
\usepackage{lmodern}
\usepackage[capitalize]{cleveref}
\usepackage{anyfontsize}
\usepackage{hyperref}
\usepackage{cleveref}

\begin{document}

\title{A Novel Kerr-like Black Hole in a General Double Power Law Dark Matter Environment: Geometry, Spectroscopy, and Energy Extraction}

\author[1]{Supakchai Ponglertsakul}
\email{supakchai.p@gmail.com}
\affil[1]{Strong Gravity Group, Department of Physics, Faculty of Science, Silpakorn University, Nakhon Pathom 73000, Thailand}

\author[2]{David Senjaya} \email{davidsenjaya@protonmail.com} 
\affil[2]{Department of Physics, Faculty of Science, Mahidol University, Bangkok 10400, Thailand}


\maketitle

\begin{abstract}

We construct a novel Kerr-like black hole solution embedded in a general double power law dark matter environment by applying the Newman--Janis algorithm to a Schwarzschild-like seed geometry. This framework provides a unified rotating spacetime for arbitrary double power law density profiles and reveals how dark matter modifies the horizon structure, extremal spin, and curvature properties of rotating black holes. Remarkably, we find that the rotation--halo interplay can eliminate essential curvature singularities for Dehnen-type profiles with $\gamma\leq2$, despite the singular nature of the corresponding static configurations. We then investigate the spectroscopic signatures of the dark matter environment through massive scalar perturbations in the Dehnen $(1,4,\gamma)$ halo. Using an analytical low-frequency matching method, we derive the quasibound state spectrum, the onset condition for scalar cloud formation, and superradiant amplification factor, showing that the halo parameters $\rho_0 r_0^3$ and $\gamma$ leave characteristic imprints on the scalar spectrum. Increasing the halo density or the cusp strengthens the binding of quasibound states and enhances their decay, shifts the scalar cloud threshold, and suppresses superradiant amplification effectivity by narrowing the allowed frequency window and lowering the amplification factor peak. Finally, we analyze rotational energy extraction from thermal scalar fields and demonstrate that the efficiency is controlled by the interplay between the thermal spectrum and the superradiant instability, with lower temperatures and less cuspy density profiles yielding more efficient energy extraction. 
\end{abstract}

\section{Introduction}

The idealized description of black holes as isolated objects is, at best, an approximation. In realistic astrophysical environments, black holes are embedded in complex and dynamically evolving systems, where interactions with surrounding matter can significantly influence their observable properties. Strong observational evidence indicates that active galactic nuclei are powered by supermassive black holes \cite{Rees1984Sep,Kormendy1995Sep}, while galaxies themselves reside within extended dark matter halos \cite{Bertone2018Oct}. These considerations naturally point toward composite systems in which black holes and dark matter coexist and interact gravitationally.

Motivated by this picture, a growing body of work has investigated black holes embedded in dark matter halos. Different halo profiles lead to distinct modifications of the spacetime geometry and its observational signatures. For instance, studies of black hole shadows, geodesic motion, and wave propagation have been carried out in Hernquist-type halos \cite{Xavier2023Mar,Cardoso2022Mar} and Dehnen-type halos \cite{Pantig:2022whj,Gohain:2024eer,Al-Badawi:2024asn,Toshmatov:2025rln,Uktamov:2025lwb,Senjaya:2025via,Senjaya:2026mkl}. Within the universal rotation curve framework, black hole solutions relevant to realistic galactic environments have been constructed and later extended to rotating configurations. Alternative approaches include piecewise halo models, general halo-dependent metrics, and rotating solutions in cold or scalar field dark matter backgrounds \cite{Konoplya2019Aug,Konoplya2022Jul,Hou2018Jul}. Optical properties, thermodynamics, gravitational lensing, and photon dynamics in such spacetimes have also been extensively explored \cite{Yang2024Jan,Liang2023Nov,Carvalho2023Dec,Anjum2023May,Pantig2022May,Stuchlik2021Nov,Pantig2023Jan,Ovgun2024Apr}.

Among the available models, the double power law density profile \cite{Dehnen1993Nov,Mo2010May} plays a particularly important role, especially in the context of dwarf galaxies. While such systems were traditionally not expected to host central black holes, recent observations suggest otherwise, with growing evidence for intermediate-mass black holes in dark matter-dominated environments \cite{BibEntry2024Jun,Bustamante-Rosell2021Nov}. This indicates that black hole--dark matter systems may be more common than previously assumed.

Understanding the interplay between dark matter halos and rotating black holes is therefore of considerable interest. The presence of dark matter can alter the spacetime geometry and lead to observable imprints in black hole shadows, accretion processes, and gravitational lensing. This is particularly relevant in light of recent horizon-scale observations, such as those of M87* and Sagittarius A* \cite{TheEventHorizonTelescopeCollaboration2019Apr,EventHorizonTelescopeCollaboration2022May}, which provide direct probes of strong gravity in realistic astrophysical environments.

Black holes also serve as natural laboratories for testing gravitational physics through gravitational waves. The first direct detection by LIGO \cite{LIGOScientific:2016aoc} opened a new observational window, enabling precise measurements of the ringdown phase governed by quasinormal modes (QNMs) \cite{Vishveshwara:1970zz,Press:1971wr,Kokkotas:1999bd}. These characteristic oscillations encode detailed information about the underlying spacetime geometry and are highly sensitive to deviations from the Kerr solution \cite{Li:2021zct}. Assessing how environmental effects, such as dark matter halos, modify the QNM spectrum is therefore essential for interpreting gravitational wave observations in realistic scenarios.

In this work, we construct a rotating black hole solution embedded in a general double power law dark matter halo by applying the Newman--Janis algorithm \cite{Newman1965Jun,Newman1965Jun1} to an analytically integrable static seed metric, and investigate its impact on scalar quasibound states and superradiant scattering. We then analyze how the presence of dark matter modifies the geometry and its associated physical properties.

Scalar perturbations provide a complementary probe of the spacetime structure and its dynamical response. Scalar perturbations obey the Klein--Gordon equation in curved spacetime, and their characteristic frequencies encode both geometric and dynamical information. In generic rotating backgrounds, the radial equation does not admit closed-form solutions, and one typically resorts to numerical or semi-analytical methods such as WKB approximations, asymptotic iteration techniques, or continued fraction expansions. While effective, these approaches often obscure the analytic structure of the spectrum.

In certain cases, however, the radial equation can be reduced to confluent or general Heun equations. Imposing the polynomial condition on these functions leads directly to discrete quantization rules for quasibound states. This analytic approach has been successfully implemented across a variety of black hole spacetimes formulated in Boyer--Lindquist coordinates \cite{Yang,100,vier21,Senjaya:2025pyv,senjaya5,Senjaya:2024uqg,Senjaya:2025bbp,Senjaya:2024gpb,Senjaya:2024blm,Senjaya:2025cgk,Senjaya:2024gfh}.

Nevertheless, in more general rotating geometries, exact analytic solutions of the full radial problem are typically not accessible. In such cases, one may instead employ the analytical asymptotic matching (AAM) method \cite{Furuhashi,Hod:2013zza,Benone:2014ssa,Huang:2016qnk}. This approach exploits the natural separation of scales in the spacetime. In the near-horizon region, where the black hole geometry dominates, the radial equation simplifies and admits solutions in terms of hypergeometric functions ${}_2F_1$, subject to purely ingoing boundary conditions at the horizon. In the asymptotic region, where the spacetime approaches flatness, the equation reduces to a confluent hypergeometric form described by ${}_1F_1$ functions.

By analytically extending both solutions into an overlapping region and matching them consistently, one can determine the relative amplitudes of ingoing and outgoing modes. This procedure yields an explicit analytic expression for the superradiant amplification factor. Such analytic control is particularly valuable for understanding how rotation and additional physical parameters influence superradiant scattering.

The remainder of this paper is organized as follows. We derive static spherically symmetric black hole with double power law dark matter halo in Sec.~\ref{sec2}. As ``seed'', the static solution will be implemented in the Newmann--Janis method to generate stationary axially symmetric solution in Sec.~\ref{secRot}. In Sec.~\ref{sec3}, we investigate the scalar perturbations, where in the Subsec.~\ref{sec4}, we study scalar quasibound states by deriving the radial solutions in the far and near regions and performing the corresponding matching procedure and Subsec.~\ref{sec5} is devoted to scalar superradiant scattering, where we obtain the amplification factor. In Sec.~\ref{sec6}, we investigate superradiant energy extraction by thermal radiation. Finally, we summarize our results and discuss their implications in Sec.~\ref{sec7}.

\section{Constructing Schwarzschild-like Black Hole with Dark Matter}\label{sec2}
To investigate the gravitational backreaction of a dark matter halo on a central compact object, we construct a static, spherically symmetric black hole solution sourced by a dark matter density profile (to be defined later). The spacetime is described by the metric ansatz
\begin{gather}
ds^2 = -f(r)\,dt^2 + \frac{dr^2}{f(r)} + r^2\left(d\theta^2 + \sin^2\theta\, d\phi^2\right),
\label{metric_ansatz}
\end{gather}
where $f(r)$ is an unknown function to be determined from the Einstein field equations.

We model the dark matter halo as an anisotropic fluid with energy--momentum tensor
\begin{equation}
T^\mu_{\ \nu} = \mathrm{diag}\left[-\rho_{\rm DM}(r),\, p_r(r),\, p_t(r),\, p_t(r)\right],
\end{equation}
where $\rho_{\rm DM}(r)$ is the dark matter density, $p_r(r)$ and $p_t(r)$ denote the radial and tangential pressures, respectively.

The Einstein field equations $G^\mu_{\ \nu} = 8\pi T^\mu_{\ \nu}$ yield
\begin{align}
G^t_{\ t} &= \frac{1}{r^2}\left[r f'(r) + f(r) - 1\right] =- 8\pi \rho_{\rm DM}(r), \\
G^r_{\ r} &= \frac{1}{r^2}\left[r f'(r) + f(r) - 1\right] = 8\pi p_r(r), \\
G^\theta_{\ \theta} &= G^\phi_{\ \phi} = \frac{1}{2} f''(r) + \frac{1}{r} f'(r) = 8\pi p_t(r),
\end{align}
where a prime denotes differentiation with respect to $r$.

An immediate consequence of the above field equations is that the matter source is not arbitrary, but is tightly constrained by the chosen metric ansatz. In particular, one observes that
\begin{equation}
G^t_{\ t} =  G^r_{\ r},
\end{equation}
which through the Einstein equations gives
\begin{equation}
p_r(r) = -\rho_{\rm DM}(r).
\end{equation}

Thus, the radial pressure is necessarily negative and equal in magnitude to the energy density. This relation is a consistency condition imposed by the spacetime geometry, implying that once the spacetime metric depends on a single function $f(r)$, the stress--energy tensor is not fully independent. Since $p_t(r) \neq p_r(r)$, the matter distribution is intrinsically anisotropic, corresponding to an effective fluid with directional stresses.

From a physical perspective, $p_r = -\rho_{\rm DM}$ represents a radial tension that balances gravitational attraction and supports a static configuration \cite{Konoplya_2026}. Consequently, the spacetime is interpreted as a black hole immersed in a galactic halo. As noted by Konoplya \cite{Konoplya_2026}, this construction naturally leads to regular and asymptotically flat black hole solutions.

Let us go back to the temporal component of the Einstein equation, one obtains a first-order differential equation for $f(r)$,
\begin{equation}
\frac{d}{dr}\left[r\left(1 -f(r)\right)\right] = 8\pi r^2 \rho_{\rm DM}(r),
\end{equation}
which can be directly integrated to give
\begin{align}
f(r) &= 1 -\frac{C}{r}- \frac{8\pi}{r}\int r'^2 \rho_{DM}
(r') dr'  \label{general_h}, \\
 &\equiv 1 -\frac{C}{r}- \frac{2 M_{DM}(r)}{r},
\end{align}
where we have defined dark matter halo mass function $M_{DM}$. The metric function is to be determined once the $\rho_{DM}$ is specified. Here in this work, we consider a general double power law  $\left(\alpha,\beta,\gamma\right)$ profile, the energy density is given by
\begin{equation}
\rho_{\rm DM}(r) = \rho_0 \left(\frac{r}{r_0}\right)^{-\gamma} \left(1 + \left(\frac{r}{r_0}\right)^\alpha\right)^{\frac{\gamma - \beta}{\alpha}}, \label{rhoDM}
\end{equation}
where $\rho_0$ and $r_0$ are the characteristic density and scale radius, respectively. The parameters $\alpha,\beta,\gamma$ determine the shape of density profile. 

Substituting this profile into Eq.~\eqref{general_h} and performing the integration analytically, we obtain
\begin{equation}
f(r) = 1 - \frac{r_s}{r} - \frac{8\pi \rho_0 r_0^\gamma}{3-\gamma }\, r^{2 - \gamma} {}_2F_1\left(\frac{3-\gamma}{\alpha},\frac{\beta-\gamma}{\alpha};1+\frac{3-\gamma}{\alpha};-\left(\frac{r}{r_0}\right)^\alpha\right),
\label{metricf}
\end{equation}
where ${}_2F_1$ is Hypergeometric function. The integration constant $r_s=2M$ is fixed such that 
the Schwarzschild solution is recovered in the case $\rho_0=0$. We remark that similar result is obtained in \cite{Cardoso2022Mar} where the authors express general halo mass function with the density profile \eqref{rhoDM}. In addition, we shall particularly focus on $0\leq\gamma\leq3$. Once the metric function is fixed the radial and tangential pressure are obtained 
\begin{align}
    p_r &= -\rho_0 \left(\frac{r}{r_0}\right)^{-\gamma} \left(1 + \left(\frac{r}{r_0}\right)^\alpha\right)^{\frac{\gamma - \beta}{\alpha}}, \\
    p_t &= \frac{\rho_0}{2}\left(\frac{r}{r_0}\right)^{-\gamma}\left(1 + \left(\frac{r}{r_0}\right)^{\alpha}\right)^{-\frac{\alpha+\beta-\gamma}{\alpha}}\left[(\gamma-2)-(\beta-2)\left(\frac{r}{r_0}\right)^\alpha\right].
\end{align}
The double power law profile \eqref{rhoDM} is very general as it can lead to several known dark matter density profiles. In Table~\ref{tab:variousf}, we derive exact black hole solution for several dark matter density profiles. More recently, observational signatures of black hole with the Dehnen type profiles are discussed in \cite{Boltaev:2026prm}.

\begin{figure}[h]
    \centering
    \includegraphics[scale=0.5]{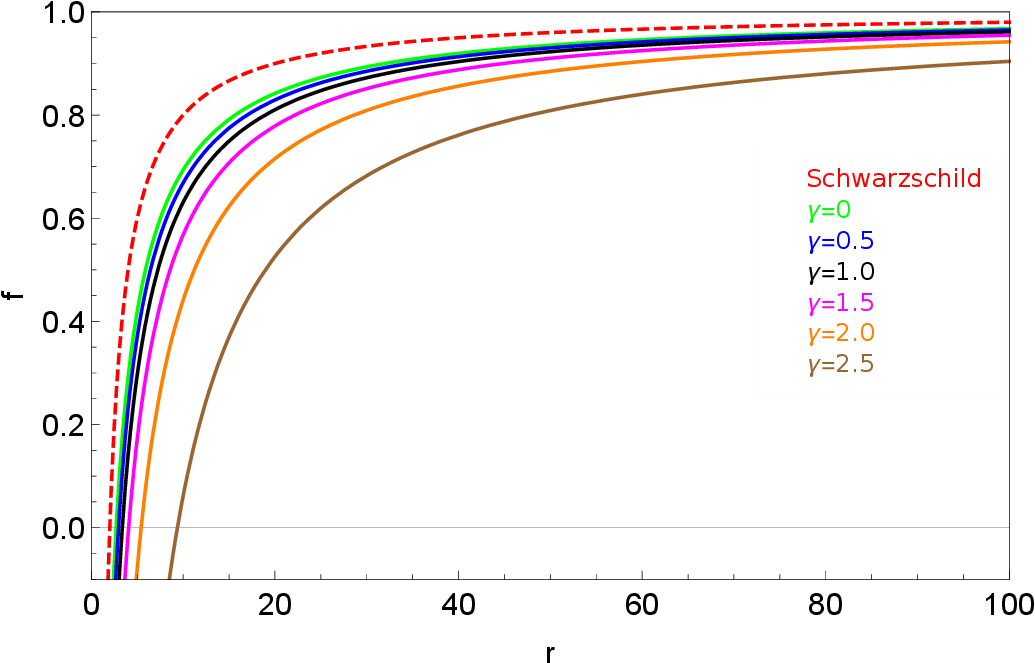}
    \caption{The metric functions $f$ for the various Dehnen profiles as a function of $r$ for $M=1,\rho_0=0.7$ and $r_0=0.6$. } \label{fig:metric}
\end{figure}

As a demonstration, we illustrate the metric function for the Dehnen type in Fig.~\ref{fig:metric}. The presence of dark matter makes black hole comparatively larger than the Schwarzschild black hole. Interestingly, as $\gamma$ increases the horizon radius (zero of $f$) becomes larger. It is obvious that the metric approaches flat spacetime at larger $r$. This nature aligns perfectly with those discussed in \cite{Boltaev:2026prm}.

\begin{table}
    \centering    
    \renewcommand{\arraystretch}{3.5}
    \begin{tabular}{|c|c|}\hline
       Profiles  &  $f(r)$ \\ \hline
       $(1,3,1)$ Navarro-Frenk-White \cite{Navarro:1995iw} & $ \displaystyle 1 - \frac{r_s}{r} + 8\pi \rho_0 r_0^3\left(\frac{1}{r+r_0} -\frac{1}{r}\ln \left[1+\frac{r}{r_0}\right]\right)$ \\ \hline
       $(1,4,1)$ Hernquist \cite{1990ApJ356359H} & $ \displaystyle 1-\frac{r_s}{r} - \frac{4\pi \rho_0 r_0^3 r}{\left(r+r_0\right)^2}$ \\ \hline
       $(1,4,2)$ Jaffe \cite{1983MNRAS202995J} & $\displaystyle 1-\frac{r_s}{r} - \frac{8\pi \rho_0 r_0^3 }{r+r_0}$\\ \hline
       $(1,4,\gamma)$ Dehnen \cite{dehnen,Boltaev:2026prm}   &  $\displaystyle 1 - \frac{r_s}{r} + \frac{8\pi \rho_0 r_0^3}{\gamma-3}\left(r+r_0\right)^{\gamma-3}r^{2-\gamma}$ \\ \hline
      $(2,4,0)$ Perfect sphere \cite{perfectsphere}  & $\displaystyle 1 - \frac{r_s}{r} + 4\pi \rho_0 r_0^3 \left( \frac{r_0}{r^2+r_0^2} - \frac{\tan^{-1}\left(r/r_0\right)}{r}\right)$ \\ \hline
      $(2,5,0)$ Plummer sphere \cite{Plummer}  & $\displaystyle 1-\frac{r_s}{r} - \frac{8\pi \rho_0 r^2}{3}\left(1 + \frac{r^2}{r_0^2}\right)^{-3/2}$  \\ \hline
    \end{tabular}
    \caption{Exact Schwarzschild-like solutions with several known density profiles.}
    \label{tab:variousf}
\end{table}

\begin{table}
    \centering
    \renewcommand{\arraystretch}{3.5}
    \begin{tabular}{|c|c|}\hline
       Profiles  &  $\displaystyle \lim_{r\to 0} \{R,R^2,K\}$ \\ \hline
       $(1,3,1)$ NFW  & $ \displaystyle \left \{\frac{24\pi \rho_0 r_0}{r} + \mathcal{O}(r^0), \frac{160\pi^2 \rho_0^2r_0^2}{r^2} + \mathcal{O}(r^{-1}), \frac{48M^2}{r^6} + \mathcal{O}(\frac{\rho_0^2}{r^2})\right\}$ \\ \hline
       $(1,4,0)$ Dehnen  & $ \displaystyle \left \{ 32\pi\rho_0 - \mathcal{O}(r), 256\pi^2\rho_0^2 - \mathcal{O}(r), \frac{48M^2}{r^6} - \mathcal{O}(\frac{\rho_0}{r^2})\right\}$ \\ \hline
       $(2,4,0)$ Perfect sphere  & $ \displaystyle \left \{ 32\pi\rho_0 - \mathcal{O}(r^2), 256\pi^2\rho_0^2 - \mathcal{O}(r^2), \frac{48M^2}{r^6} - \mathcal{O}(\frac{\rho_0}{r})\right\}$ \\ \hline
      $(2,5,0)$ Plummer sphere  & $ \displaystyle \left \{ 32\pi\rho_0 - \mathcal{O}(r^2), 256\pi^2\rho_0^2 - \mathcal{O}(r^2), \frac{48M^2}{r^6} - \mathcal{O}(\frac{\rho_0}{r})\right\}$  \\ \hline
    \end{tabular}
  \caption{Three scalar curvature invariants, the Ricci scalar $R$, the Ricci square $R^2=R_{\mu\nu}R^{\mu\nu}$ and the Kretschmann scalar $K=R_{\mu\nu\sigma\rho}R^{\mu\nu\sigma\rho}$ for various dark matter halo. }
    \label{tab:scalarinv}
\end{table}

Now, we analyze behaviour of scalar invariants of these exact solutions. We consider the Ricci scalar $R$, the Ricci squared $R^2=R_{\mu\nu}R_{\mu\nu}$ and the Kretschmann scalar $K=R_{\mu\nu\sigma\rho}R^{\mu\nu\sigma\rho}$. It turns out that these scalar invariants for general halo profile i.e., arbitrary $(\alpha,\beta,\gamma)$, are very complicated. Therefore as an example, we list three scalar curvatures in the limit $r\to0$ for the NFW, the Dehnen $(1,4,0)$, the perfect sphere and the Plummer sphere in Table~\ref{tab:scalarinv}. It is clear that without dark matter $(\rho_0=0)$, $R$ and $R^2$ vanish identically while $K=\frac{48M^2}{r^6}$. This resembles with the Schwarzschild case. In the NFW case, all scalar invariants diverge as $r\to 0$. In contrast, the profiles with $\gamma=0$, both $R,R^2$ display regularity while the Kretschmann scalar remains diverge as $r\to 0$. In addition, the $\gamma=0$ profiles show similar behaviour at the leading order while the difference appears in the sub-leading order. These clearly demonstrate that the exact solutions considered in this table possess singular nature of the spacetimes. We also find that all three curvature scalars vanish asymptotically
\begin{align}
    \lim_{r\to \infty} \{R,R^2,K\} &\to 0,
\end{align}
for all the profiles we have explored. It is worth mentioning that the behaviour of three principal curvature invariants for the Dehnen halo $(1,4,\gamma)$ is investigated in \cite{Boltaev:2026prm}.

\section{Constructing General Double Power Law Rotating Black Hole}\label{secRot}

Astrophysical black holes are expected to carry angular momentum, making rotation an essential ingredient of any realistic spacetime description. A convenient and widely adopted way to introduce rotation into a static geometry is provided by the Newman--Janis algorithm (NJA) \cite{Newman1965Jun,Newman1965Jun1}. This method generates a stationary, axisymmetric solution from a spherically symmetric seed metric through a complex transformation of the coordinates. Originally developed to obtain the Kerr solution from the Schwarzschild solution \cite{Drake1997Jul}, it has since proven effective in a broad range of gravitational contexts \cite{Brauer2015Jan,Lombardo2004Feb,Kim2025Jan,Abbas2024Apr,Alexeyev2025Mar,Jafarzade2025Jun,Fazzini2025Feb,Li2025Jan,Fathi2025Mar,Zahid2025Feb,Raza2025Jan}.

In what follows, we apply the NJA to construct the rotating counterpart of the Schwarzschild--general double power law spacetime. We begin with the static line element 
\begin{equation}
ds^2 = -F(r)\,dt^2 + \frac{dr^2}{G(r)} + H(r)\left(d\theta^2 + \sin^2\theta\,d\phi^2\right),
\label{metric1}
\end{equation}
and introduce advanced Eddington--Finkelstein coordinates $(u,r,\theta,\phi)$ via
\begin{equation}
dt = du + \frac{dr}{\sqrt{F(r)G(r)}}.
\label{trans_eq}
\end{equation}
Thus the static line element in null coordinates is
\begin{align}
    ds^2 &= - F(r) du^2 - \frac{2\sqrt{F(r)G(r)}}{G(r)}dudr + H(r)^2\left(d\theta^2+\sin^2\theta d\phi^2\right).
\end{align}
The metric is then rewritten in terms of a null tetrad basis $(l^\mu,n^\mu,m^\mu,\bar{m}^\mu)$,
\begin{equation}
g^{\mu\nu} = -l^{\mu}n^{\nu} - l^{\nu}n^{\mu} + m^{\mu}\bar{m}^{\nu} + m^{\nu}\bar{m}^{\mu}, \label{inversemetric}
\end{equation}
with
\begin{equation}
\begin{aligned}
l^{\mu} &= \delta^{\mu}_{r}, \qquad
n^{\mu} = \delta^{\mu}_{u} - \frac{1}{2}F(r)\,\delta^{\mu}_{r}, \\
m^{\mu} &= \frac{1}{\sqrt{2H(r)}}\left(\delta^{\mu}_{\theta} + \frac{i}{\sin\theta}\delta^{\mu}_{\phi}\right),
\end{aligned}
\end{equation}
where $\bar{m}^{\mu}$ denotes the complex conjugate of $m^\mu$. These vectors satisfy the following conditions
\begin{align}
    l_\mu l^\mu = n_\mu n^\mu = m_\mu m^\mu = \bar{m}_\mu \bar{m}^\mu = l_\mu m^\mu = n_\mu m^\mu = 0, \nonumber\\
    l_\mu n^\mu = 1,~~~~~~ m_\mu \bar{m}^\mu = -1.
\end{align}
We note that for the case of our interest \eqref{metric_ansatz}, $F(r)=G(r)=f(r)$ and $H(r)=r^2$.

The key step of the NJA is a complex transformation of the coordinates,
\begin{equation}
u \rightarrow u' = u - ia\cos\theta, 
\qquad 
r \rightarrow r' = r + ia\cos\theta,
\end{equation}
where $a$ is a rotation parameter. This transformation effectively introduces rotation by mixing the radial and angular coordinates of the metric. After the transformation, the new null tetrad vectors \cite{Bambi:2013ufa} are
\begin{align}
    l'^\mu = \delta^\mu_r,~~n'^\mu = \delta^\mu_u - \frac{F(r')}{2}\delta^\mu_r,~~
    m'^\mu = \frac{1}{\sqrt{2H(r')}}\left(ia\sin\theta\left(\delta^\mu_u-\delta^\mu_r\right) + \delta^\mu_\theta + \frac{i \delta^\mu_\phi}{\sin\theta}\right). \label{newtetrad}
\end{align}
Therefore, one can construct an inverse metric tensor corresponding to the new tetrad above via \eqref{inversemetric}. The non-vanishing component of inverse metric tensor based on the null tetrad \eqref{newtetrad} are 
\begin{align}
    g^{uu} = \frac{a^2\sin^2\theta}{H(r,\theta)},~~~~~g^{ur} = -1-\frac{a^2\sin^2\theta}{H(r,\theta)},~~~~~g^{u\phi} = \frac{a}{H(r,\theta)}, \nonumber \\
    g^{rr} = F(r,\theta) + \frac{a^2\sin^2\theta}{H(r,\theta)},~~~~~g^{r\phi} = -\frac{a}{H(r,\theta)}, \nonumber \\
    g^{\theta\theta} = \frac{1}{H(r,\theta)},~~~~~g^{\phi\phi} = \frac{1}{H(r,\theta)\sin^2\theta}, \label{inverseNew}
\end{align}
where we denote $\displaystyle \lim_{a \to0} F(r,\theta),H(r,\theta) \to f(r),r^2$. To put the new metric above to the Boyer-Lindquist coordinates $(t,r,\theta,\phi)$, we must eliminate $du$ and the cross term $drd\phi$. This can be done by the following coordinate transformation
\begin{align}
    du &= dt + \lambda(r)dr,~~~~~d\phi = d\varphi + \chi(r)dr,
\end{align}
where $\lambda$ and $\chi$ solely depend on $r$. In the Boyer-Lindquist coordinates, there is only one cross term, $dtdr$. This condition fixes $\lambda$ and $\chi$ to be \cite{Bambi:2013ufa}
\begin{align}
    \lambda(r) &= -\frac{H(r,\theta)+a^2\sin^2\theta}{F(r,\theta)H(r,\theta)+a^2\sin^2\theta},~~~~~\chi(r) = -\frac{a}{F(r,\theta)H(r,\theta)+a^2\sin^2\theta}.
\end{align}
We obtain the rotating solution in the Boyer-Lindquist coordinates
\begin{align}
    ds^2 &= -F(r,\theta)dt^2 - a(1-F(r,\theta))\sin^2\theta dt d\phi + \frac{dr^2}{F(r,\theta)+\frac{a^2\sin^2\theta}{H(r,\theta)}} \nonumber \\
    &~~~~+ H(r,\theta)d\theta^2 + \sin^2\theta\left[H(r,\theta)+(2-F(r,\theta))a^2\sin^2\theta\right]d\phi^2. \label{dsrotate}
\end{align}
Following the prescription of \cite{Azreg-Ainou2014Sep,Yang:2022uze}, we promote the metric functions according to
\begin{align}
    F(r,\theta) &\to \frac{f(r)r^2+a^2\cos^2\theta}{\left(r^2+a^2\cos^2\theta\right)^2}\rho^2, \\
    H(r,\theta) &\to \rho^2 \equiv r^2+a^2\cos^2\theta.
\end{align}
With these transformations, it is easy to show that $\lambda$ and $\chi$ are function of radial coordinate $r$. Now, the spacetime metric \eqref{dsrotate} can be put into the Kerr-like form
\begin{align}
ds^2 &= -\left(1 -\frac{\xi(r)}{\rho^2}\right)dt^2 
+ \frac{\rho^2}{\Delta}dr^2 + \rho^2 d\theta^2 - \frac{2\xi(r)}{\rho^2}a\sin^2\theta\, dt\, d\varphi \nonumber \\
&~~~~
+\frac{\Sigma(r,\theta)}{\rho^2}\sin^2\theta\, d\varphi^2, \label{Kerrlike}
\end{align}
where
\begin{align}
\xi(r) &= r^2\left(1 - f(r)\right), \label{xi}\\
\Delta(r) &= r^2+a^2-\xi(r),\\
\Sigma(r,\theta) &= (r^2+a^2)^2 -\Delta a^2 \sin^2\theta.
\end{align}
This is the rotating counterpart of the static seed metric \eqref{metricf} obtained within the chosen NJA prescription. Setting $a=0$ leads to static spherically symmetric solution \eqref{metric_ansatz}. Moreover, it reduces to the Schwarzschild and the Kerr black hole in an appropriate limit. It is important to mention that it has been proved in \cite{Azreg_A_nou_2014} that the obtained solution above satisfies the Einstein equation. 

\begin{figure}[h]
    \centering
    \includegraphics[scale=0.35]{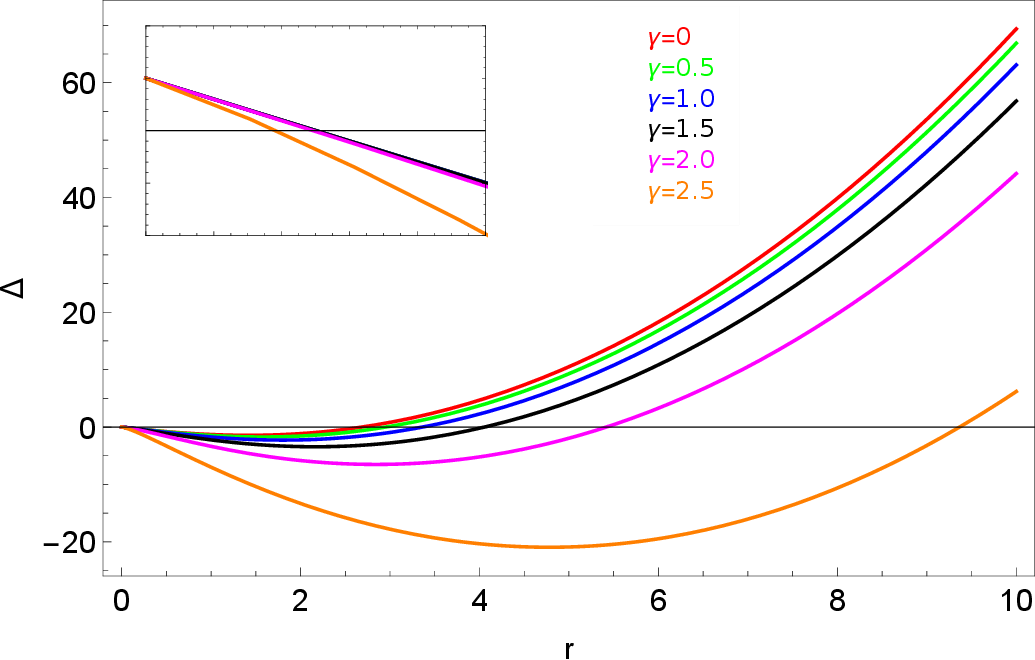}
    \includegraphics[scale=0.35]{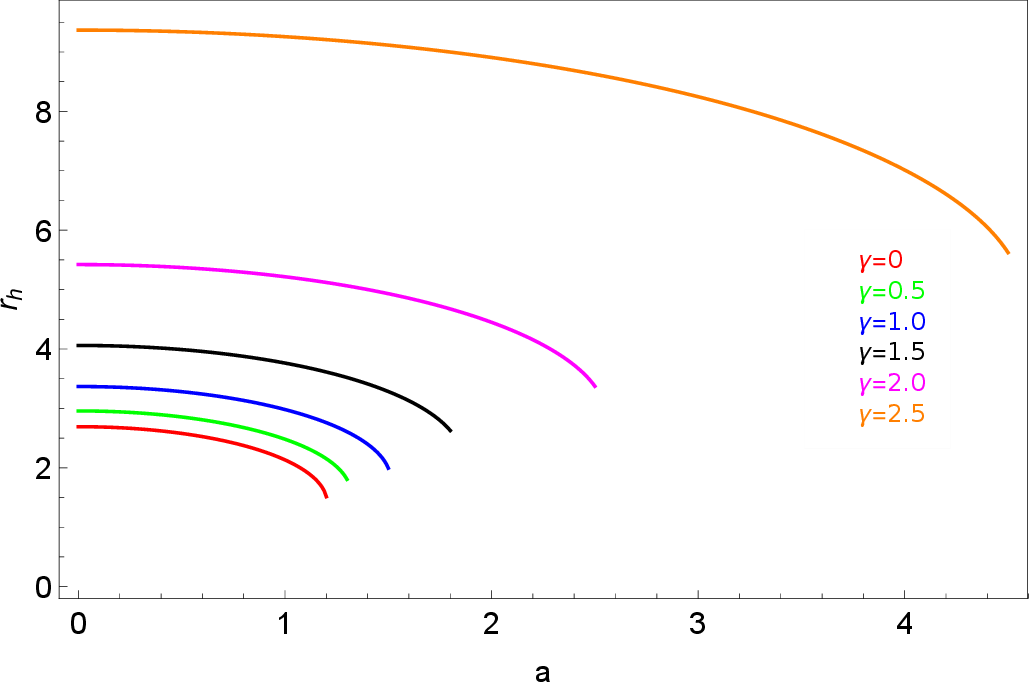}
    \caption{Left: the behaviour of $\Delta$ against radial coordinate $r$ for various Dehnen profiles $(1,4,\gamma)$, $M=1, \rho_0=0.7,r_0=0.6$ and $a=0.4$. The sub-plot displays a close up behaviour at small $r$. Right: The outer event horizon for various Dehnen profiles against $a$.} \label{fig:metricrot}
\end{figure}

Since the seed metric $f(r)$ is asymptotically flat, therefore, it is easy to show that the rotating solution above is also asymptotically flat. The locations of horizons are determined from $\Delta(r_h)=0.$ Unfortunately, there is no closed-form without specify ($\alpha,\beta,\gamma$). Therefore, the horizon will be explored numerically. This is demonstrated in Fig.~\ref{fig:metricrot}. For illustration purpose, here, the dark matter halo is described by the general Dehnen profile ($1,4,\gamma$) with fixed $\rho_0=0.7$ and $r_0=0.6$. It is clear that the rotating solution possesses two horizons, inner and outer horizon. For given $a=0.4$, the outer horizon increases as $\gamma$ increases. In the sub-figure, we notice the location of inner horizon. Interestingly, we also observe that the $\Delta$ function is regular at $r=0$ as can be seen from the sub-figure. The right figure displays the behaviour of the outer horizon $r_h$ as a function of rotation parameter $a$. In general, the horizon radii decrease with $a$. We find that larger values of $\gamma$ allow the black hole to attain higher spins before reaching extremality.

\begin{figure}[h]
    \centering
    \includegraphics[scale=0.35]{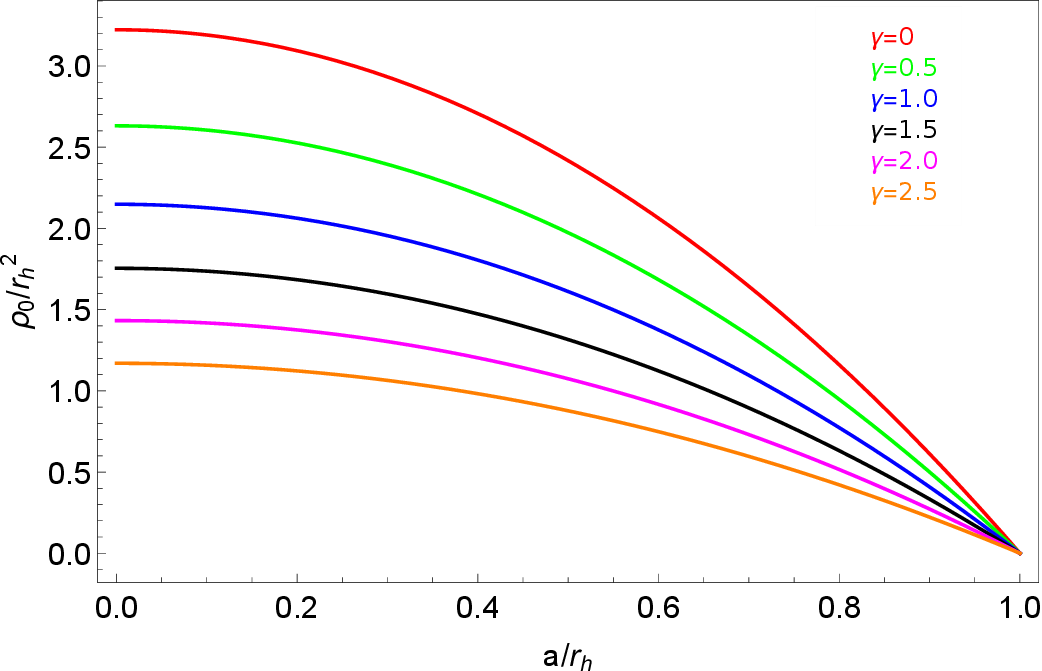}
    \includegraphics[scale=0.35]{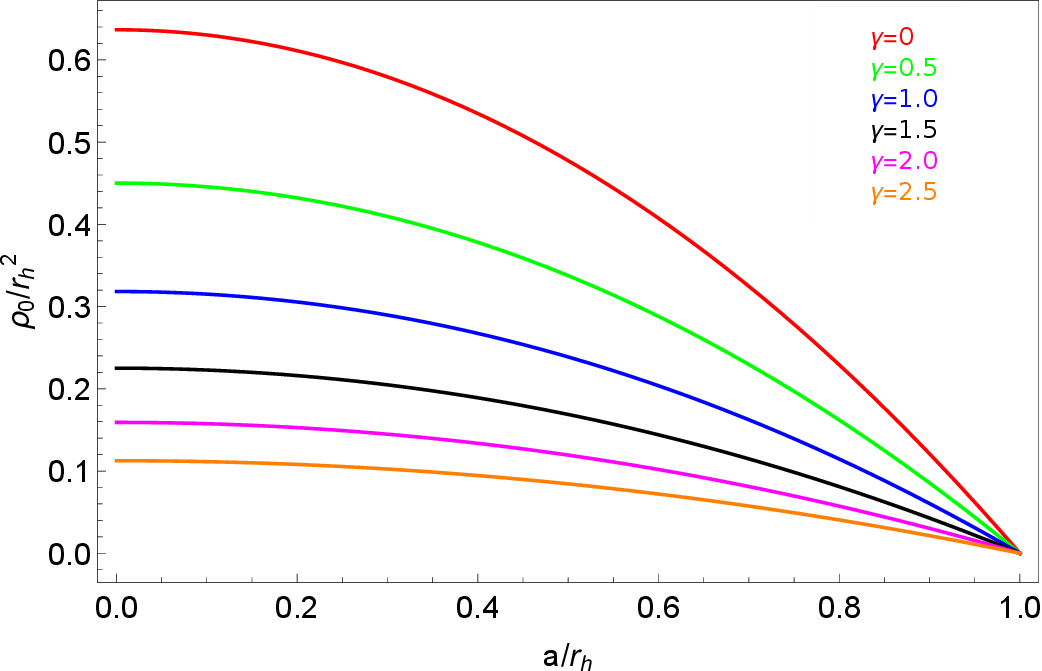}
    \caption{Parameter space of the black hole (regions under the curves), with extremal lines indicated by the colored curves. Here, we fix $r_0/r_h$ = 0.5 (left) and 1.0 (right).} \label{fig:extremal}
\end{figure}

Moreover, we illustrate the parameter space in which the rotating black hole has two horizons, represented by the region under the colored curves in Fig.~\ref{fig:extremal}. The boundary between the region below and above the curves denotes the extremal limit, where the black hole’s two horizons coincide. We observe the allowed parameter region decreases as the inner slope parameter $\gamma$ increases. In addition, as the dimensionless length scale $r_0/r_h$ increases, the parameter space generally decreases.

\begin{figure}[h]
    \centering
    \includegraphics[scale=0.23]{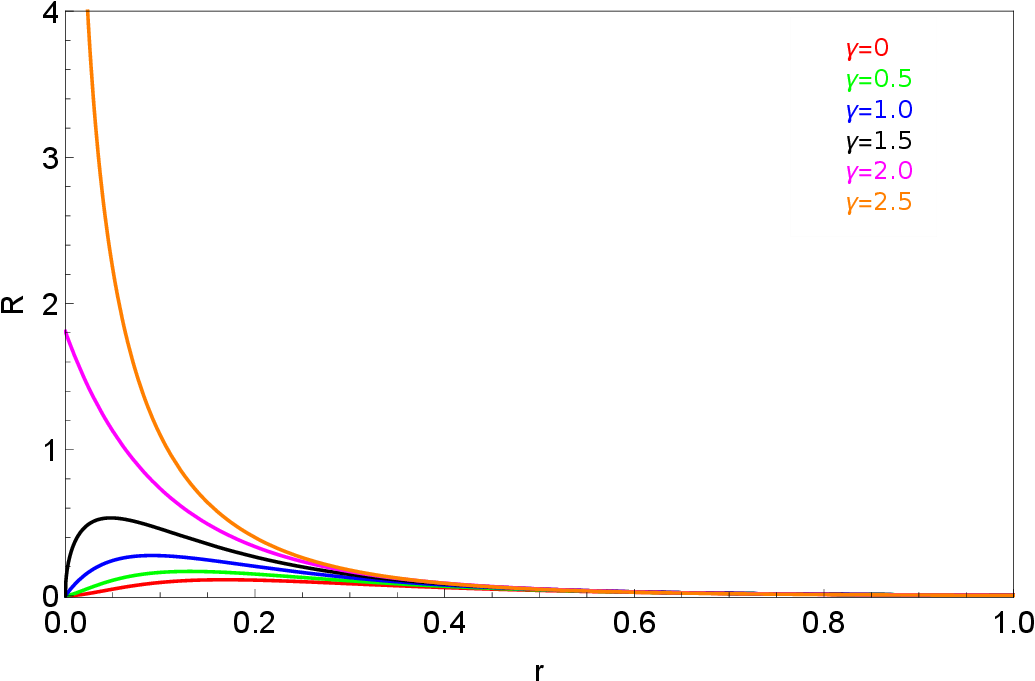}
    \includegraphics[scale=0.23]{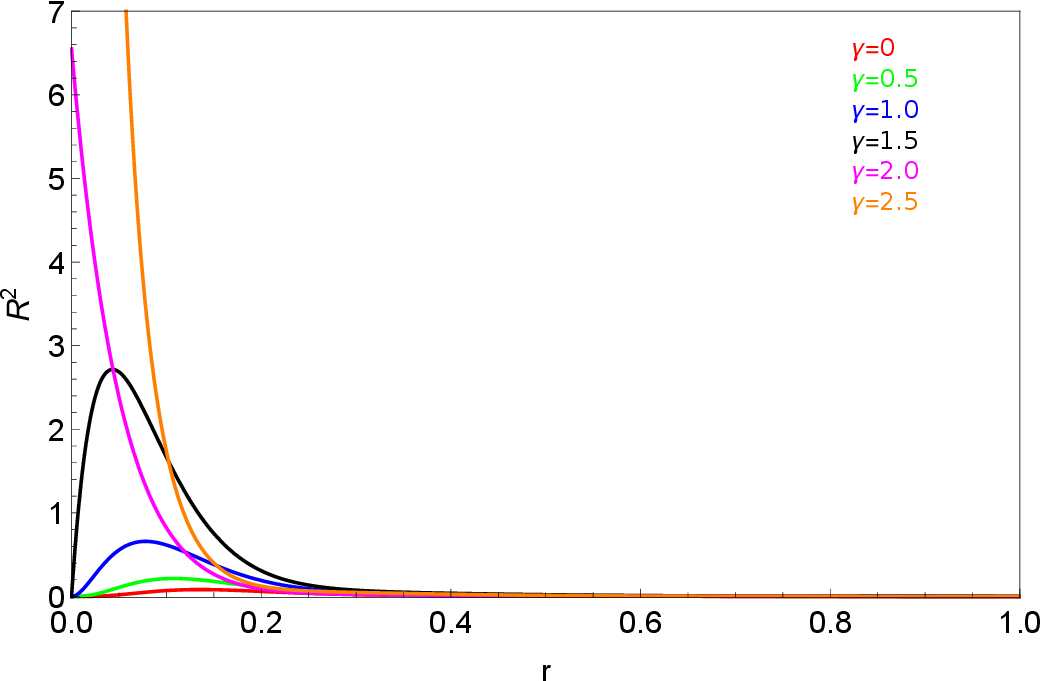}
    \includegraphics[scale=0.23]{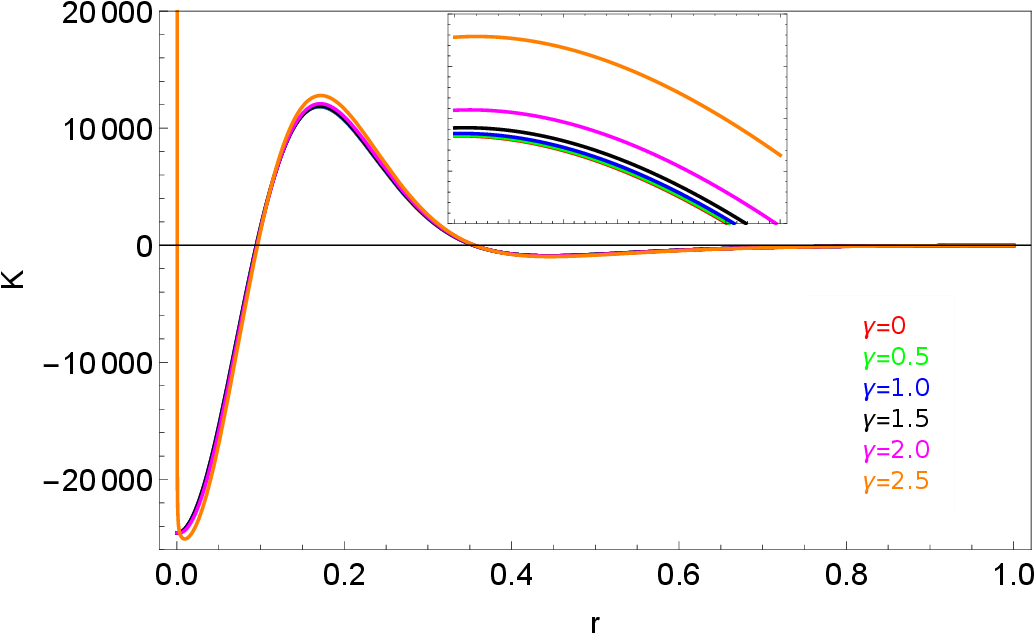}
    \caption{Three curvature scalars for $M=1,a=0.5,\theta=\pi/4,\rho_0=0.1$ and $r_0=0.3$ for various $\gamma$ profiles. In the middle panel, Ricci tensor squared for $\gamma\leq 3/2$ is scaled by a factor of 5 for illustrative purpose. In the right panel, an inset is included to make the difference more visible.} \label{fig:curveinvt}
\end{figure}

Next, we investigate curvature singularity of the solution \eqref{Kerrlike}. We explore three curvature scalars, $R,R^2,K$. For the Dehnen profile $(1,4,\gamma)$. these quantities can be shown schematically as 
\begin{align}
    \lim_{\theta \to \theta \neq \frac{\pi}{2}}\left(\lim_{r\to0} R \right) &\propto r^{2-\gamma}, \\
    \lim_{\theta \to \theta \neq \frac{\pi}{2}}\left(\lim_{r\to0} R^2 \right) &\propto r^{4-2\gamma}, \\
    \lim_{\theta \to \theta \neq \frac{\pi}{2}}\left(\lim_{r\to0} K \right) &\propto \{r^{0},r^{-1}\}.
\end{align}
It is found that the Kretschmann scalar is proportional to constant for $\gamma \leq 2$ and to $r^{-1}$ for $\gamma > 2$. From these expressions, it is observed that the three curvature invariants are regular for $\gamma \leq 2$ as $r\to 0$. Thus, the rotating black hole with Dehnen dark matter profile $(\gamma \leq 2)$ has no essential singularity. This is surprising, since the static seed solution \eqref{metricf} possesses essential singularity regardless of the value of $\gamma$ \cite{Boltaev:2026prm}. The behaviour of three invariant curvatures are illustrated in Fig.~\ref{fig:curveinvt}. It is clear that all curvature scalars are regular at small $r$ for $\gamma\leq 2$. In contrast, for $\gamma=5/2$, they are singular at small $r$.

\begin{figure}[h]
    \centering
    \includegraphics[scale=0.23]{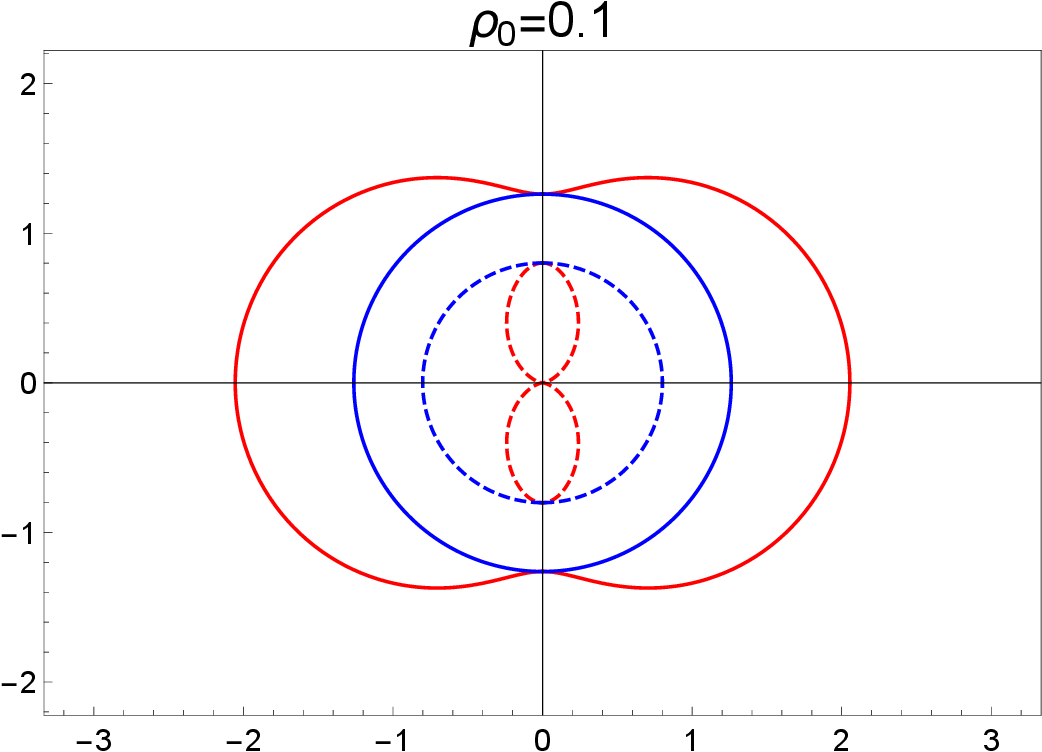}
    \includegraphics[scale=0.23]{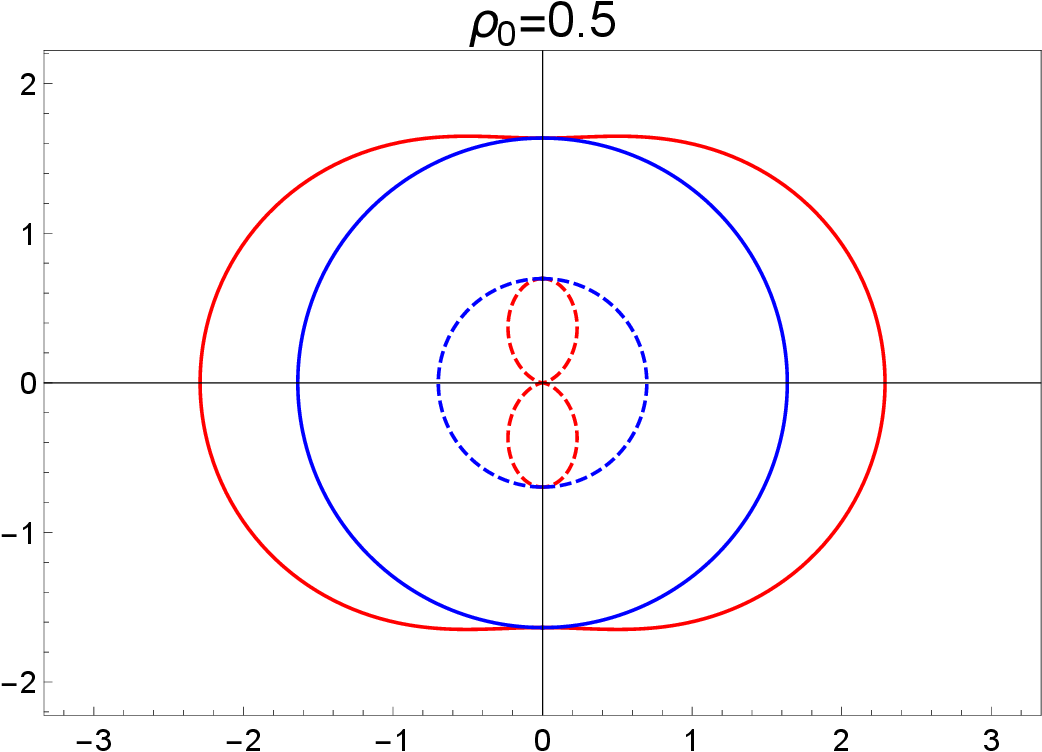}
    \includegraphics[scale=0.23]{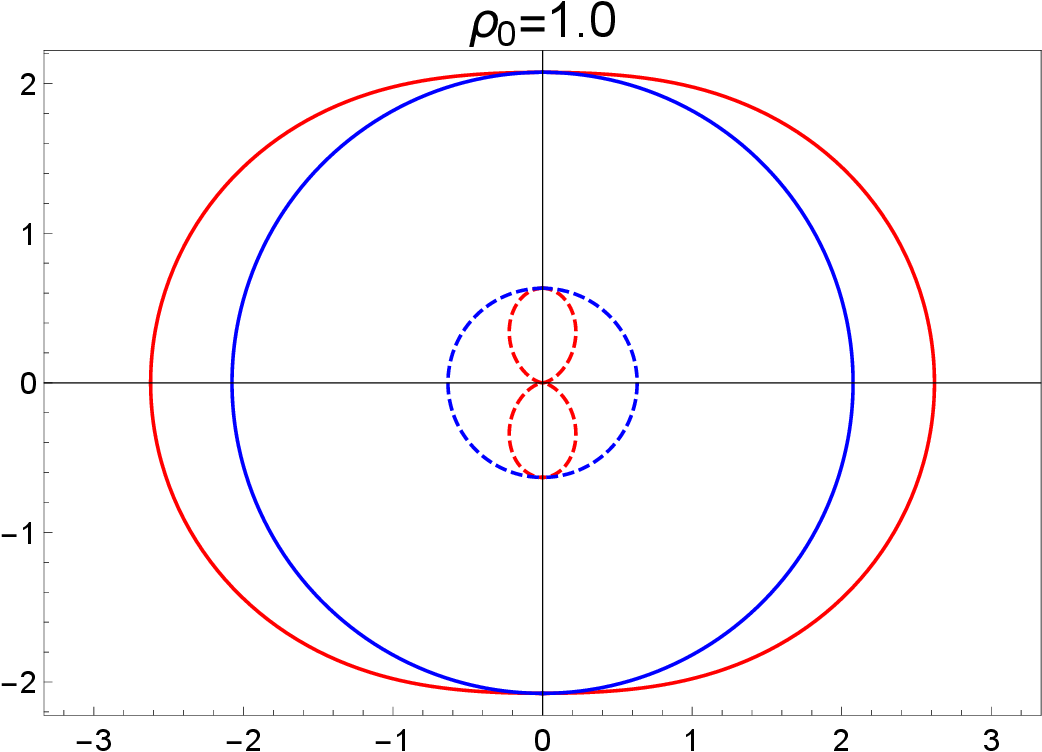}
    \caption{Parametric plots of the outer and inner stationary-limit surfaces ( $r_{s_\pm}$ shown in red) and the outer and inner event horizons (shown in blue) for the cored-type dark matter profile with $\gamma=0$. In these plots, we fix $M=1,r_0=0.5$ and $a=0.99$, and consider three distinct values of $\rho_0.$} \label{fig:ergo}
\end{figure}

For a rotating black hole, there exists a region in which observers cannot remain static. The boundary separating the region in which observers cannot remain at rest from the region where they can remain static is called the stationary-limit surface. Such surface can be determined numerically from 
\begin{align}
\left(1 -\frac{\xi(r)}{\rho^2}\right) &= 0.    
\end{align}
In turns out that, there are two positive real roots for the above representing outer $r_{s_+}$ and inner $r_{s_{-}}$ stationary-limit surfaces. In Fig.~\ref{fig:ergo}, we display stationary-limit surfaces of rotating black hole surrounded by cored-type Dehnen dark matter i.e., $\gamma=0$. In this figure, $r_{s_+}$ and $r_{s_-}$ are denoted by the red solid and the red dashed curves where the outer and inner horizon are given in blue color. 

We find that varying the Dehnen profile parameter $\gamma$ has only a mild effect on the stationary-limit surface. Therefore, as an illustrative example, we consider only the case $\gamma=0$ here. We observe that stationary-limit surfaces and the event horizon of black hole coincide at poles $\theta=0,\pi/2.$ A clear general trend can be observed: the sizes of the outer stationary-limit surface and outer event horizon increase with the central density $\rho_0$, whereas those of the inner surfaces decrease as $\rho_0$ increases. In addition, the concave shape at the pole becomes less apparent as $\rho_0$ gets bigger. 

\section{Scalar Perturbation} \label{sec3}
We now consider a scalar field propagating in curved spacetime. The Klein--Gordon equation in natural units ($\hbar = c =G = 1$) reads
\begin{equation}
\frac{1}{\sqrt{-g}} \partial_\mu \left(\sqrt{-g}\, g^{\mu\nu} \partial_\nu \psi \right) - m^2 \psi = 0,
\label{KG_explicit}
\end{equation}
where $m$ is the scalar field mass and $\sqrt{-g}=\rho^2\sin\theta$. 

To proceed, we exploit the stationarity and axial symmetry of the spacetime and introduce the separation ansatz
\begin{equation}
\psi(t,r,\theta,\phi)=e^{-i\omega t + i m_\ell \phi} R(r)\,\Theta(\theta).
\end{equation}
Substituting this ansatz into \eqref{KG_explicit} and dividing by the full wavefunction, we obtain
\begin{multline}
\frac{1}{R}\frac{d}{dr}\left(\Delta \frac{dR}{dr}\right)
+ \frac{1}{\Theta \sin\theta}\frac{d}{d\theta}\left(\sin\theta \frac{d\Theta}{d\theta}\right) \\
+ \frac{(r^2+a^2)^2 \omega^2 - 2a m_\ell \omega (r^2+a^2-\Delta) + a^2 m_\ell^2}{\Delta}
\\- a^2 \omega^2 \sin^2\theta- \frac{m_\ell^2}{\sin^2\theta}
-m^2 r^2 - a^2 m^2 \cos^2\theta=0.
\label{separated}
\end{multline}

Rearranging the terms into purely radial and angular contributions yields
\begin{multline}
\frac{1}{R}\frac{d}{dr}\left(\Delta \frac{dR}{dr}\right)
+ \frac{(r^2+a^2)^2 \omega^2 - 2a m_\ell \omega (r^2+a^2-\Delta) + a^2 m_\ell^2}{\Delta}
- m^2 r^2 -a^2\omega^2 \\
= -\left[
\frac{1}{\Theta \sin\theta}\frac{d}{d\theta}\left(\sin\theta \frac{d\Theta}{d\theta}\right)
+ a^2(\omega^2 - m^2)\cos^2\theta
- \frac{m_\ell^2}{\sin^2\theta}
\right].
\end{multline}

Since the left-hand side depends only on $r$ and the right-hand side only on $\theta$, both sides must be equal to a separation constant, denoted by $\lambda_\ell^{m_\ell}$.

The angular equation becomes
\begin{equation}
\frac{1}{\sin\theta}\frac{d}{d\theta}\left(\sin\theta \frac{d\Theta}{d\theta}\right)
+\left[a^2(\omega^2-m^2)\cos^2\theta - \frac{m_\ell^2}{\sin^2\theta} + \lambda_\ell^{m_\ell} \right]\Theta = 0,
\end{equation}
whose solutions are the spheroidal harmonics and the radial equation is obtained as
\begin{equation}
\frac{d}{dr}\left(\Delta \frac{dR}{dr}\right) +\left[ \frac{\big((r^2+a^2)\omega - a m_\ell\big)^2}{\Delta} - m^2 r^2 -a^2\omega^2+2 a m_\ell \omega - \lambda_\ell^{m_\ell} \right] R = 0.
\label{radial_final_clean}
\end{equation}
Now, we shall derive analytic expression of quasibound state in rotating spacetime with the Dehnen profiles.

\subsection{Quasibound State} \label{sec4}
To obtain analytic solutions of the radial equation~\eqref{radial_final_clean}, we employ the analytical asymptotic matching method (AAM). This approach is particularly effective in the regime
\begin{equation}
m M \ll 1, \qquad |\omega| M \ll 1,
\end{equation}
that is, both the scalar field mass and the frequency are small and of the same order. We further assume that the rotation parameter satisfies
\begin{equation}
m a \ll 1, \qquad |\omega| a \ll 1.
\end{equation}
Under these conditions, the angular equation reduces to that of spherical harmonics, and the separation constant simplifies to
\begin{equation}
\lambda_\ell^{m_\ell} \approx \ell(\ell+1), \qquad \ell = 0,1,2,\dots
\end{equation}
We now focus on the radial equation~\eqref{radial_final_clean}. The core idea of the asymptotic matching method is to solve the equation in two distinct regions:
\begin{itemize}
\item the near-horizon region, 
\item the far region,
\end{itemize}
and then match the two solutions in their overlapping domain.

\subsubsection{Far-region solution}
We first consider the far region defined by
\begin{equation}
r \gg M,
\end{equation}
which also implies $r \gg a$. 

To determine the asymptotic behavior of the metric functions $\Delta,f$ at large distances, we introduce the parameters
\begin{equation}
a_1=\frac{3-\gamma}{\alpha},
\qquad
b_1=\frac{\beta-\gamma}{\alpha},
\qquad
c_1=1+\frac{3-\gamma}{\alpha}=a_1+1 .
\end{equation}
The seed function can then be written in the compact form
\begin{equation}
f(r) = 1-\frac{r_s}{r} + \frac{8\pi\rho_0 r_0^\gamma}{\gamma-3}\, r^{2-\gamma} \,{}_2F_1 \!\left( a,b;c; -\left(\frac{r}{r_0}\right)^\alpha \right).
\end{equation}
The asymptotic expansion of the Gauss hypergeometric function for large $|z|$ is given by (see Appendix \ref{AppendixA}) 
\begin{equation}
{}_2F_1(a_1,b_1;c_1;|z|\to\infty) \approx \frac{\Gamma(c_1)\Gamma(b_1-a_1)} {\Gamma(b_1)\Gamma(c_1-a_1)} (-z)^{-a_1} + \frac{\Gamma(c_1)\Gamma(a_1-b_1)} {\Gamma(a_1)\Gamma(c_1-b_1)} (-z)^{-b_1},
\end{equation}

Substituting these expressions into $f(r)$, the large-$r$ behavior becomes
\begin{align}
f(r) &\approx  1 -\frac{r_s}{r} - \frac{8\pi\rho_0 r_0^3}{3-\gamma} \frac{\Gamma(c_1)\Gamma(b_1-a_1)} {\Gamma(b_1)\Gamma(c_1-a_1)} \frac{1}{r} -\frac{8\pi\rho_0 r_0^\beta}{3-\gamma} \frac{\Gamma(c_1)\Gamma(a_1-b_1)} {\Gamma(a_1)\Gamma(c_1-b_1)} \,r^{2-\beta}, \\
&\equiv 1 + \frac{A}{r} + \frac{B}{r^{\beta-2}}.
\end{align}

For completeness, it is useful to discuss several commonly used values of the outer slope parameter $\beta$. Since the asymptotic correction behaves as $r^{2-\beta}$, different choices of $\beta$ lead to qualitatively distinct asymptotic structures.
\begin{itemize}
\item {$\beta=2$ (Modified Isothermal Sphere).}
In this case, the next-to-leading contribution becomes a constant,
\begin{equation}
f(r) \approx 1+B +\frac{A}{r}.
\end{equation}

Since the density decreases as $\rho\sim r^{-2}$, the enclosed mass grows linearly with radius,
$M(r)\propto r$. Consequently, the total mass diverges and the spacetime is not asymptotically Schwarzschild.

\item {$\beta=3$ (NFW and Moore Profiles).}
For $\beta=3$, the asymptotic correction scales as $1/r$,
\begin{equation}
f(r) \approx 1+\frac{A+B}{r}.
\end{equation}

The additional contribution merely renormalizes the coefficient of the Schwarzschild term. Although the mass divergence is weaker than for $\beta=2$, the total mass still diverges logarithmically,
$M(r)\propto \ln r$, and therefore remains formally infinite.

\item  {$\beta=4$ (Hernquist, Jaffe, and Dehnen Family Profiles).}
A particularly interesting case arises for $\beta=4$, for which the next-to-leading contribution scales as $r^{-2}$. The asymptotic expansion then takes the form
\begin{equation}
f(r) \approx 1+\frac{A}{r} +\frac{B}{r^2} . \label{fexpand}
\end{equation}

In contrast to the previous cases, the density falls off sufficiently rapidly, $\rho\sim r^{-4}$, ensuring that the total mass converges. As a result, the spacetime possesses a well-defined asymptotically Schwarzschild behavior with a genuine $r^{-2}$ correction, analogous to the charge term in the Reissner--Nordström solution.
\end{itemize}

We now specialize to the case $\alpha=1$ and $\beta=4$, i.e., the Dehnen family profiles. In this case, the asymptotic expansion of the metric function $f(r)$ is given by
\begin{equation}
f(r) \approx 1-\frac{r_s}{r}
-\frac{8\pi\rho_0 r_0^3}{(3-\gamma)\,r} + \mathcal{O}\left(\frac{1}{r^2}\right). \label{fapprox}
\end{equation}
Thus, this identifies
\begin{equation}
A=-r_s-\frac{8\pi\rho_0 r_0^3}{3-\gamma}.
\end{equation}
The metric function is approximated as $\Delta \approx r^2+ A r$. In far region limit, the radial equation~\eqref{radial_final_clean} simplifies to
\begin{equation}
r^2 \frac{d}{dr}\left(r^2 \frac{dR}{dr}\right)
+\left[(\omega^2 - m^2)r^4 - \lambda_\ell^{m_\ell} r^2 + m^2 r^2 \xi(r)\right]R = 0.
\label{far_approx_1}
\end{equation}
By substituting approximation \eqref{fapprox} into \eqref{xi} and \eqref{far_approx_1}, we obtain
\begin{equation}
\frac{d^2}{dr^2}(rR)
+\left[\omega^2 - m^2 - \frac{Am^2}{r} - \frac{\ell(\ell+1)}{r^2}\right] rR = 0.
\label{far_approx_2}
\end{equation}
To bring this equation into a standard form, we define
\begin{equation}
k^2 \equiv m^2 - \omega^2, \qquad 
\nu \equiv -\frac{Am^2}{2k}, \qquad 
x \equiv 2kr, \label{knu}
\end{equation}
then \eqref{far_approx_2} becomes
\begin{equation}
\frac{d^2}{dx^2}(xR)
+\left[-\frac{1}{4} + \frac{\nu}{x} - \frac{\ell(\ell+1)}{x^2} \right] xR = 0.
\label{far_approx_3}
\end{equation}
This is the standard confluent hypergeometric equation (see Appendix \ref{AppendixB}). Imposing the boundary condition of a decaying (bound-state) solution at infinity and regular at the horizon, the solution reads
\begin{equation}
R_\infty(x) = C_1\, x^{\ell} e^{-x/2} {}_1F_1(\ell+1-\nu,\,2\ell+2,\,x),
\label{far_solution}
\end{equation}
where $C_1$ is a constant.

It is instructive to note that this solution reduces to the hydrogen-like wavefunction when $-(\ell+1-\nu)\equiv n_r$ is an integer and acts as the radial quantum number. However, in the present case, the frequency $\omega$ generally acquires a small imaginary part due to the interaction with the rotating black hole. Consequently, $\nu$ is also complex
\begin{equation}
\nu = \ell + 1 + n_r + \delta\nu, \label{quantization}
\end{equation}
where $n_r$ is a positive integer and $\delta\nu$ is the small imaginary correction. 

The far-region solution can then be rewritten as
\begin{equation}
R_\infty(x) = C_1\, x^{\ell} e^{-x/2} {}_1F_1(-n_r-\delta\nu,\,2\ell+2,\,x).
\label{far_solution_final}
\end{equation}

We now determine the asymptotic form of the far-region solution~\eqref{far_solution_final} in the near-horizon regime. The overlapping region between the far and near-horizon domains is defined by
\begin{equation}
M \ll r \ll \max\left(\frac{\ell}{m}, \frac{\ell}{|\omega|}\right),
\end{equation}
which is consistent with the assumptions $m M \ll 1$ and $|\omega|M \ll 1$. In this region, the condition
\begin{equation}
|x| = 2|k|r \ll 1
\end{equation}
holds.

Expanding the confluent hypergeometric function in \eqref{far_solution_final} in the small-$x$ limit (see \eqref{1F1_small_x_eps} in Appendix \ref{AppendixB}) 
\begin{multline}
\lim_{r\rightarrow 0} R_\infty(r) \approx C_1 (-1)^{n_r} \frac{(2\ell+1+n_r)!}{(2\ell+1)!} (2k)^{\ell} r^{\ell}
\\+ C_1 (-1)^{n_r+1} (2\ell)! \,n_r! \, \delta\nu \, (2k)^{-\ell-1} r^{-\ell-1}.
\label{far_small_clean}
\end{multline}

\subsubsection{Near-horizon solution}
We now turn to the near-horizon region and solve the radial equation in this limit. Expanding the metric function near the outer horizon $r_h$, we write
\begin{equation}
\xi(r) \approx \xi(r_h) + \xi'(r_h)(r-r_h),
\end{equation}
that the function $\Delta(r)$ takes the leading-order form
\begin{equation}
\Delta(r) \approx r^2+a^2-\left[\xi(r_h) + \xi'(r_h)(r-r_h)\right], \label{11}
\end{equation}
where
\begin{equation}
    \Delta(r_h)=0=r_h^2+a^2-\xi(r_h) \quad \to \quad \xi(r_h)=r_h^2+a^2.
\end{equation}

Substituting the identity above into \eqref{11} yields
\begin{equation}
    \Delta(r) \approx (r-r_h)\bigl(r -r_h + 2\kappa\bigr),
\end{equation}
where we have introduced,
\begin{equation}
\kappa \equiv r_h - \frac{1}{2}\xi'(r_h).
\end{equation}

Let us introduce the following the dimensionless coordinate
\begin{equation}
z=\frac{r-r_h}{2\kappa},
\end{equation}
and defining
\begin{equation}
P=-\frac{\omega \xi(r_h)-a m_\ell}{2\kappa}.
\end{equation}

We further consider the low-frequency, $\{\omega,m\} \ll 1$, for which $1/\omega \gg 1$. In this limit, the scalar-field frequency $\omega$ plays a crucial role in determining the size of the overlap region where the asymptotic matching approximation (AAM) is valid. The near-region condition is therefore defined as $\omega z \ll 1$, that implies $r-r_h \ll \frac{r_h}{\omega}$, whereas the far-region requirement, $z \gg 1$, corresponds to $r_h \ll r-r_h$. Combining these inequalities yields the matching region
\begin{equation}
r_h \ll r-r_h \ll \frac{r_h}{\omega},
\end{equation}
within which both the near- and far-region solutions are simultaneously valid. It follows that the size of the overlap region scales as $1/\omega$; hence, as $\omega$ decreases, the matching region becomes increasingly broad, leading to a more accurate asymptotic matching procedure. In particular, the condition $\omega \ll 1$ guarantees the existence of a parametrically large overlap region.

Within the near-region limit $\omega z \ll 1$, the radial equation simplifies to
\begin{equation}
z(z+1)\frac{d}{dz}\left[z(z+1)\frac{dR}{dz}\right]
+\left[P^2 - \lambda_\ell^{m_\ell} z(z+1)\right]R = 0.
\end{equation}
To bring this equation into a standard form, we introduce the transformation
\begin{equation}
R(z)=\left(\frac{z}{z+1}\right)^{iP} Y(z),
\end{equation}
which yields the hypergeometric equation
\begin{equation}
z(z+1)Y'' + (1+2iP+2z)Y' - \lambda_\ell^{m_\ell} Y = 0.
\end{equation}

The general solution is therefore given by (see Appendix \ref{AppendixC})
\begin{multline}
Y(z) = A_1\, {}_2F_1(-\ell,\ell+1,1+2iP;-z) \\
+ A_2\, (-z)^{-2iP} \, {}_2F_1(-\ell-2iP,\ell+1-2iP,1-2iP;-z).
\end{multline}

The physical boundary condition at the event horizon is defined by requiring that the solution be purely ingoing as seen by an observer co-rotating with the black hole. In this frame, no outgoing flux can emerge from the horizon, which selects the physically admissible mode. Imposing this condition in the near-horizon limit ($z \to 0$) selects the solution
\begin{equation}
r_h(z)=A_1 \left(\frac{z}{z+1}\right)^{iP}
{}_2F_1(-\ell,\ell+1,1+2iP;-z),
\label{near_solution_clean}
\end{equation}
where the exponent $iP$ corresponds to the ingoing wave behavior at the horizon in the co-rotating frame. 

We now extract the near horizon's asymptotic behavior in the large-$r$ limit ($z \gg 1$) (see \eqref{GaussConnection} in Appendix \ref{AppendixA}). This yields
\begin{multline}
\lim_{r\rightarrow \infty} r_h(r) \approx  A_1 \frac{\Gamma(1+2iP)\Gamma(2\ell+1)}{\Gamma(\ell+1)\Gamma(\ell+1+2iP)} (2\kappa)^{-\ell} r^\ell\\ + A_1 \frac{\Gamma(1+2iP)\Gamma(-2\ell-1)}{\Gamma(-\ell)\Gamma(2iP-\ell)} (2\kappa)^{\ell+1} r^{-\ell-1}.
\label{near_large_clean}
\end{multline}
The far- and near-region solutions exhibit the same leading asymptotic behavior in the overlap region. The matching is therefore performed by equating the coefficients of the $r^\ell$ and $r^{-\ell-1}$ modes in \eqref{far_small_clean} and~\eqref{near_large_clean}, yielding
\begin{equation}
C_1 \bar \alpha = A_1\bar  \gamma,
\qquad
C_1 \bar \beta\, \delta\nu = A_1 \bar \delta.
\end{equation}
where
\begin{align}
\bar \alpha &= (-1)^{n_r} \frac{(2l+1+n_r)!}{(2\ell+1)!}(2k)^\ell, \\
\bar \beta  &= (-1)^{n_r+1}(2\ell)! \, n_r! \,(2k)^{-\ell-1}, \\
\bar \gamma &= \frac{\Gamma(1+2iP)\Gamma(2\ell+1)}{\Gamma(\ell+1)\Gamma(\ell+1+2iP)} (2\kappa)^{-\ell}, \\
\bar \delta &= \frac{\Gamma(1+2iP)\Gamma(-2\ell-1)}{\Gamma(-\ell)\Gamma(2iP-\ell)} (2\kappa)^{\ell+1}.
\end{align}
Eliminating the normalization constants $C_1$ and $A_1$ by taking the ratio of the two equations, we obtain
\begin{equation}
\delta\nu = \frac{\bar{\alpha}}{\bar{\beta}} \cdot \frac{\bar{\delta}}{\bar{\gamma}}.
\end{equation}

We now evaluate each factor explicitly. First,
\begin{equation}
\frac{\bar \alpha}{\bar \beta}=-\frac{(2\ell+1+n_r)!}{(2\ell+1)!(2\ell)!n_r!}(2k)^{2\ell+1}.
\end{equation}

Next, for the near-horizon contribution, we find
\begin{equation}
\frac{\bar \delta}{\bar \gamma}=
\frac{\Gamma(\ell+1)}{\Gamma(2\ell+1)}
\frac{\Gamma(-2\ell-1)}{\Gamma(-\ell)}
\frac{\Gamma(\ell+1+2iP)}{\Gamma(2iP-\ell)}(2\kappa)^{2\ell+1}.
\end{equation}
Using Gamma-function identities \cite{bell1968special}, these are, in particular,
\begin{align}
\frac{\Gamma(\ell+1+2iP)}{\Gamma(2iP-\ell)}&= (2iP)(-1)^\ell \prod_{j=1}^{\ell}(j^2+4P^2), \\
\frac{\Gamma(-2\ell-1)}{\Gamma(-\ell)}&= (-1)^{\ell+1}\frac{\ell!}{(2\ell+1)!}.
\end{align}
Together with $\Gamma(\ell+1)=\ell!$, we obtain
\begin{equation}
\frac{\bar{\delta}}{\bar{\gamma}}=-2iP\,(2\kappa)^{2\ell+1}\frac{(\ell!)^2}{(2\ell)!(2\ell+1)!}\prod_{j=1}^{\ell}(j^2+4P^2).
\end{equation}
Combining the above results, we finally arrive at
\begin{equation}
\delta\nu=2iP\,(4k\kappa)^{2\ell+1}\frac{(2\ell+1+n_r)!}{n_r!}\left[\frac{\ell!}{(2\ell)!(2\ell+1)!}\right]^2\prod_{j=1}^{\ell}(j^2+4P^2). \label{dnu}
\end{equation}
Using the definitions of $k$, $\nu$, and the quantization condition \eqref{knu} and \eqref{quantization}, one obtains 
\begin{equation}
m^2-\omega^2=\frac{A^2m^4}{4(n+\delta\nu)^2},
\label{spectrum_eq}
\end{equation}
where $n=n_r+\ell+1$ is the principal quantum number. Since $\delta\nu$ represents a small correction to the hydrogenic spectrum, equation \eqref{spectrum_eq} can be solved perturbatively by expanding the frequency as
\begin{equation}
\omega=\omega^{(0)}+\delta\nu\, \omega^{(1)}+\mathcal{O}(\delta\nu^2). \label{a}
\end{equation}
At leading order, neglecting $\delta\nu$ in \eqref{spectrum_eq} yields
\begin{equation}
m^2-\left(\omega^{(0)}\right)^2=\frac{A^2m^4}{4n^2}.
\end{equation}
Solving for the frequency, we obtain 
\begin{equation}
\omega^{(0)}=m\sqrt{1-\frac{A^2m^2}{4n^2}}\approx m\left(1-\frac{A^2m^2}{8n^2}\right),
\label{omega0}
\end{equation}
which reproduces the familiar hydrogen-like spectrum. Therefore, this implies
\begin{equation}
k=\sqrt{m^2-\left(\omega^{(0)}\right)^2}\approx\frac{|A|m^2}{2n}.
\label{kleading}
\end{equation}
Furthermore, since $\omega^{(0)}\approx m$, we may approximate 
\begin{equation}
P = -\frac{\omega\xi(r_h)-a m_\ell}{2\kappa} \approx -\frac{m\xi(r_h)-a m_\ell}{2\kappa}. 
\label{Pleading}
\end{equation}
Substituting \eqref{kleading} and \eqref{Pleading} into \eqref{dnu}, we obtain
\begin{multline}
\delta\nu=-2i\left[\frac{m\xi(r_h)-a m_\ell}{2\kappa}\right] \left(\frac{2|A|\kappa m^2}{n}\right)^{2\ell+1} \times\\\frac{(2\ell+1+n_r)!}{n_r!} \left[\frac{\ell!}{(2\ell)!(2\ell+1)!}\right]^2 \prod_{j=1}^{\ell} \left(j^2+4P^2\right). \label{dnu_final}
\end{multline}
The next-to-leading-order correction is obtained by substituting
\begin{equation}
\omega=\omega^{(0)}+\delta\nu \,\omega^{(1)}
\end{equation}
into \eqref{spectrum_eq} and retaining terms linear in $\delta\nu$. This yields 
\begin{equation}
\delta\nu\,\omega^{(1)}= \frac{A^2m^3}{4n^3}\,\delta\nu.
\label{omega1}
\end{equation}
Substituting \eqref{dnu_final} into \eqref{omega1}, we find
\begin{multline}
\delta\nu\,\omega^{(1)}=-2i\frac{A^2m^3}{4n^3}\left[\frac{m\xi(r_h)-a m_\ell}{2\kappa}\right] \left(\frac{2|A|\kappa m^2}{n}\right)^{2\ell+1} \times\\\frac{(2\ell+1+n_r)!}{n_r!} \left[\frac{\ell!}{(2\ell)!(2\ell+1)!}\right]^2 \prod_{j=1}^{\ell} \left(j^2+4P^2\right).
\label{omega1_final}
\end{multline}
Combining the leading- and next-to-leading-order contributions, the quasibound state frequency is given by
\begin{multline}
\omega\approx m\left(1-\frac{A^2m^2}{8n^2}\right)-2i\frac{A^2m^3}{4n^3}\left[\frac{m\xi(r_h)-a m_\ell}{2\kappa}\right] \left(\frac{2|A|\kappa m^2}{n}\right)^{2\ell+1} \times\\\frac{(2\ell+1+n_r)!}{n_r!} \left[\frac{\ell!}{(2\ell)!(2\ell+1)!}\right]^2 \prod_{j=1}^{\ell} \left(j^2+4P^2\right).
\label{omega_final}
\end{multline}
Note that, in the massless case $(m=0)$, no potential barrier exists to support a bound state. Therefore, the above formula becomes trivial in this limit. 

Since $\delta\nu \,\omega^{(1)}$ is purely imaginary, it determines the growth or decay rate of the quasibound state. From \eqref{Pleading}, it follows that
\begin{equation}
\text{Im}(\omega)=\omega_I \propto P \propto -\left[m\xi(r_h)-a m_\ell\right].
\end{equation}
Therefore, the quasibound states become unstable whenever
\begin{equation}
m < \frac{m_\ell
a}{r_h^2+a^2}\label{instability_condition}.
\end{equation}
This is known as the black hole bomb. The black hole bomb mechanism arises when superradiant amplification is combined with a confining mechanism that prevents radiation from escaping to infinity. For Reissner-Nordstr\"om (RN) black hole, a mirror cavity \cite{Herdeiro:2013pia,Dolan:2015dha} and asymptotic anti de-Sitter (AdS) \cite{Uchikata:2011zz} are found to trigger superradiant instability. In the case of rotating spacetime, similar to RN black hole, reflecting boundary and AdS boundary render black hole bomb \cite{Cardoso:2004nk,Cardoso:2004hs}. Moreover, a presence of massive bosonic field also leads to the instability on Kerr spacetime \cite{Damour:1976kh}.

In the quasibound state regime, the field is localized outside the black hole by an effective potential barrier, forming a trapped mode. As the mode interacts with the horizon, it undergoes superradiant amplification, while the outer barrier prevents it from dispersing to infinity. The wave is therefore confined between the horizon and the potential barrier, repeatedly undergoing amplification. This sets up a self-reinforcing cycle in which energy is continuously extracted from the black hole’s rotation and fed back into the mode. As a result, the amplitude grows exponentially over time. In this way, the quasibound state acts as a feedback mechanism, steadily draining rotational energy and driving the instability.

\begin{figure}[h]
    \centering
    \includegraphics[scale=0.35]{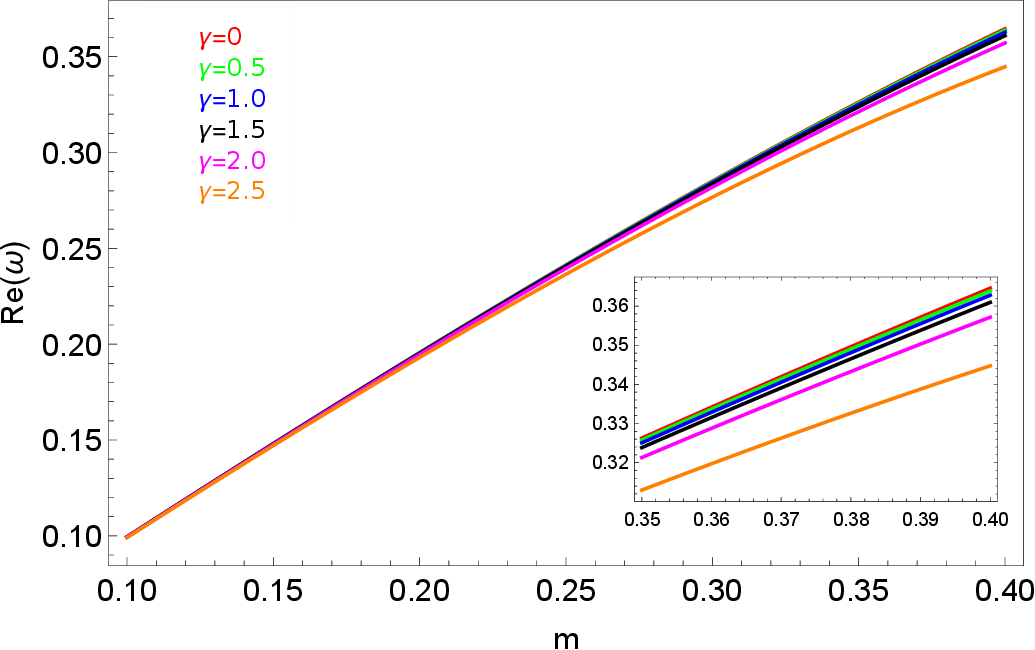}
    \includegraphics[scale=0.35]{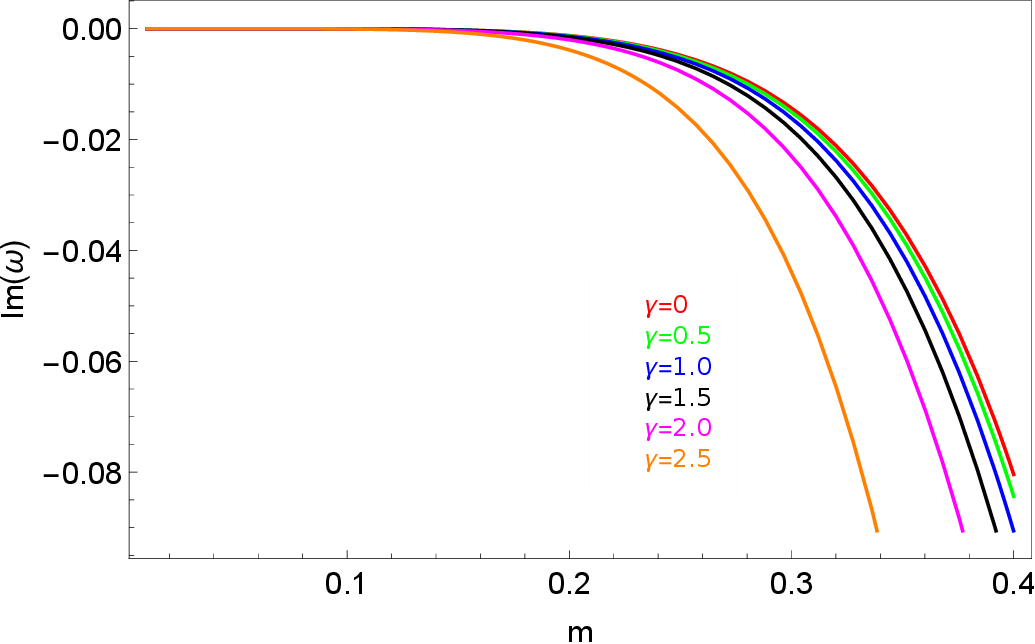}\\
    \includegraphics[scale=0.35]{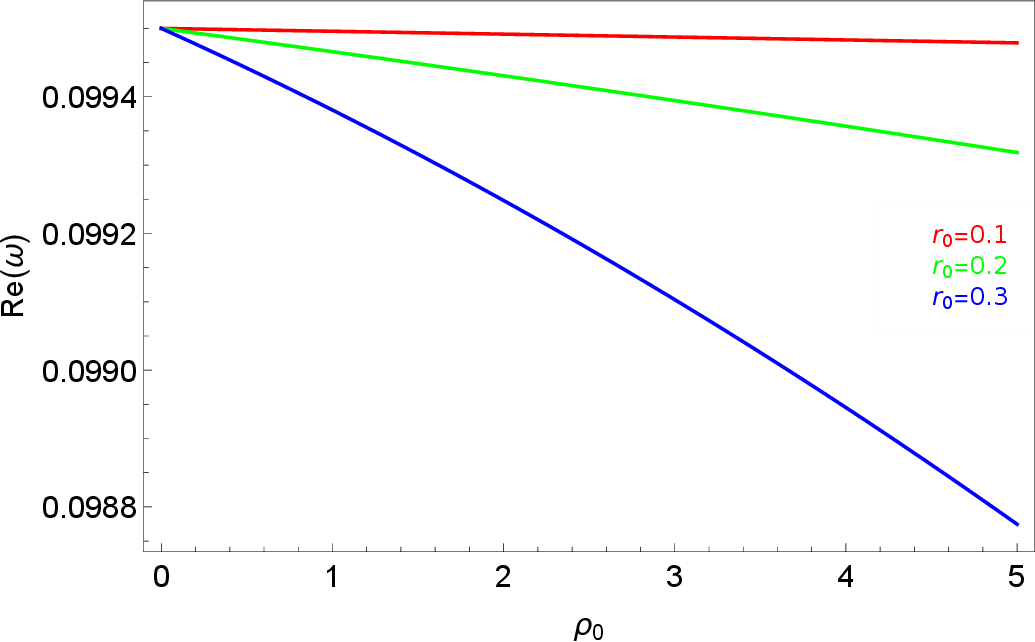}
    \includegraphics[scale=0.35]{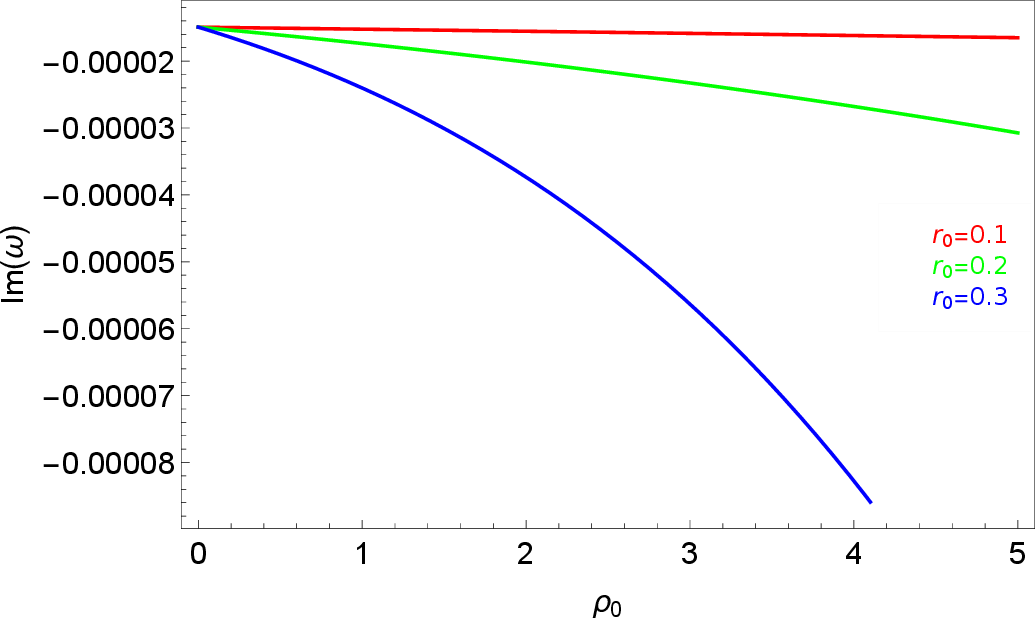}
    \caption{Real and imaginary part of the quasibound state frequencies as function of $m$ (top) and $\rho_0$ (bottom). The other parameters are fixed as (top): $n=1,\ell=0,m_\ell=0,a=0.1,\rho_0=0.1, r_0=0$. (bottom): $m=0.1,n=1,\ell=0,m_\ell=0,\gamma=0,a=0.5$. The subplot displays the behaviour at large $m\sim0.35-0.4.$ } \label{fig:qbs}
\end{figure}

Figure~\ref{fig:qbs} illustrates the dependence of the quasibound state frequencies on the scalar-field mass $m$ (top panels) and the central dark matter density $\rho_0$ (bottom panels). The real part of the frequency initially increases with $m$, reaches a maximum, and subsequently decreases as the field mass becomes larger. As the Dehnen profile parameter $\gamma$ increases, the peak shifts toward lower values of $\mathrm{Re}(\omega)$, corresponding to an increase in the binding energy $m-\mathrm{Re}(\omega)$. Simultaneously, the imaginary part becomes more negative, indicating that steeper dark matter density profiles lead to faster decay and consequently shorter-lived quasibound states.

The lower panels show the influence of the halo density parameter $\rho_0$ for different characteristic halo radii $r_0$. Increasing $\rho_0$ decreases the real part of the frequency while making the imaginary part more negative. These effects become increasingly pronounced for larger values of $r_0$, demonstrating that the halo influence is primarily governed by the effective parameter combination $\rho_0r_0^3$. Consequently, increasing either the halo density or the characteristic halo size produces more tightly bound quasibound states with shorter lifetimes.

\subsection{Superradiance Scattering}\label{sec5}
Superradiant scattering constitutes one of the most distinctive features of rotating black hole spacetimes, whereby incident bosonic waves can extract rotational energy from the black hole and emerge with an enhanced amplitude. In contrast to the quasibound state regime, which probes long-lived trapped configurations, scattering processes provide direct insight into the dynamical response of the spacetime to external perturbations. In this section, we investigate the scattering of a massive scalar field in the general double power law rotating black hole, with particular emphasis on how the presence of a dark matter halo modifies the standard Kerr superradiant behavior. To this end, we employ the analytical asymptotic matching method (AAM) to obtain an analytic expression for the amplification factor in the low-frequency regime. For scattering states, the scalar field is unbound, implying that the frequency satisfies $\omega > m$. Moreover, the frequency $\omega$ is real, as appropriate for scattering processes.

\subsubsection{Far-region solution}
Let us rewrite the far-region equation \eqref{far_approx_2}, this is 
\begin{equation}\label{5}
\frac{\mathrm{d}^2}{\mathrm{d}y^2}\left[yR(y)\right]
+\left[-\frac{1}{4}+\frac{i\Lambda}{y}-\frac{\ell(\ell+1)}{y^2}\right]yR(y)=0,
\end{equation}
where we have defined $K \equiv ik$, together with
\begin{eqnarray}
K^2 &\equiv& \omega^2 - m^2 > 0, \label{Def_3}\\
\Lambda &\equiv& \frac{A m^2}{2K}, \label{Def_4}\\
y &\equiv& 2iKr.
\end{eqnarray}

The general solution of \eqref{5} can be expressed in terms of confluent hypergeometric $ {}_1F_1$ functions as
\begin{multline}
R_\infty(y) = B_1 e^{-y/2} y^\ell \, {}_1F_1\left(\ell+1-i\Lambda,\,2\ell+2,\,y\right) \\+ B_2 e^{-y/2} y^{-\ell-1} \, {}_1F_1\left(-\ell-i\Lambda,\,-2\ell,\,y\right),\label{R(y)}
\end{multline}
where $B_1$ and $B_2$ are constants to be determined via matching the solution with the near-horizon solution. In the small-$y$ limit, the radial function behaves as 
\begin{equation}
\lim_{y\to 0} R_\infty(y) = B_1 y^{\ell} + B_2 y^{-\ell-1}.
\label{far_small_2}
\end{equation}
At large distances, we use the asymptotic expansion of the confluent hypergeometric function,
\begin{equation}
\lim_{y\to \infty} {}_1F_1(a,b,y) \approx \frac{\Gamma(b)}{\Gamma(a)} e^{y} y^{a-b}
+ \frac{\Gamma(b)}{\Gamma(b-a)} (-y)^{-a}. \label{connection1f1}
\end{equation}
Substituting this expansion into \eqref{R(y)}, each term generates contributions proportional to $e^{\pm y/2}$. Using $y=2iKr$, we identify
\begin{equation}
e^{\pm y/2} = e^{\pm i K r}
\end{equation}
which allows the asymptotic solution to be expressed in terms of the radial coordinate $r$ as
\begin{multline}
    B_1 e^{-y/2} y^\ell \, {}_1F_1\left(\ell+1-i\Lambda,\,2\ell+2,\,y\right)\approx B_1 e^{-iKr} (2iKr)^\ell\times\\\left[\frac{\Gamma(2\ell+2)}{\Gamma(\ell+1-i\Lambda)}e^{2iKr}y(2iKr)^{-\ell-1-i\Lambda}+ \frac{\Gamma(2\ell+2)}{\Gamma(\ell+1+i\Lambda)}(-2iKr)^{-\ell-1+i\Lambda}\right]
\end{multline}
and
{\begin{multline}
    B_2 e^{-y/2} y^{-\ell-1} \, {}_1F_1\left(-\ell-i\Lambda,\,-2\ell,\,y\right)\approx B_2 e^{-iKr} (2iKr)^{-\ell-1}\times\\\left[\frac{\Gamma(-2\ell)}{\Gamma(-\ell-i\Lambda)}e^{2iKr}(2iKr)^{\ell-i\Lambda}+ \frac{\Gamma(-2\ell)}{\Gamma(-\ell+i\Lambda)}(-2iKr)^{\ell+i\Lambda}\right]
\end{multline}}
regrouping the same exponent yields
\begin{multline}
\lim_{r\to \infty} R_\infty(r) \approx \left[ B_1 \frac{\Gamma(2\ell+2)}{\Gamma(\ell+1-i\Lambda)} + B_2 \frac{\Gamma(-2\ell)}{\Gamma(-\ell-i\Lambda)} \right] {\left(2iK\right)^{-i\Lambda-1}} e^{+iKr} r^{-i\Lambda-1} \\ + \left[ B_1 \frac{{\left(-1\right)^{-\ell-1+i\Lambda}}\Gamma(2\ell+2)}{\Gamma(\ell+1+i\Lambda)} + B_2 \frac{{\left(-1\right)^{\ell+i\Lambda}}\Gamma(-2\ell)}{\Gamma(-\ell+i\Lambda)} \right]{\left(2iK\right)^{-1+i\Lambda}} e^{-iKr} r^{i\Lambda-1}. \label{infinity}
\end{multline}
These terms correspond to outgoing and ingoing waves, respectively. 

\subsubsection{Near-horizon solution}
To complete the matching procedure, we now relate the far-region solution to the near-horizon behavior. Recall that the near-horizon solution takes the form
\begin{equation}
r_h(z)=A_1 \left(\frac{z}{z+1}\right)^{iP} {}_2F_1(-\ell,\ell+1,1+2iP;-z),
\label{near_solution_recall}
\end{equation}
where $z=(r-r_h)/(2\kappa)$.

In the overlap region $z\gg 1$ (equivalently $r\gg r_h$, while still within the matching domain), the hypergeometric function admits the asymptotic expansion
\begin{multline}
\lim_{r \to \infty} R_h(r) \approx 
A_1 \frac{\Gamma(1+2iP)\Gamma(2\ell+1)}{\Gamma(\ell+1)\Gamma(\ell+1+2iP)} (2\kappa)^{-\ell} r^\ell
\\
+ A_1 \frac{\Gamma(1+2iP)\Gamma(-2\ell-1)}{\Gamma(-\ell)\Gamma(2iP-\ell)} (2\kappa)^{\ell+1} r^{-\ell-1}.
\label{near_large_rewrite}
\end{multline}
This expression exhibits two independent power-law behaviors, $r^\ell$ and $r^{-\ell-1}$, which must be matched to the small-$y$ expansion of the far-region solution in \eqref{far_small_2}, namely
\begin{equation}
\lim_{r\to 0} R_\infty(r)=B_1 (2iK r)^{\ell}+ B_2 (2iK r)^{-\ell-1}.
\end{equation}
Matching coefficients of equal powers of $r$ yields
\begin{align}
B_1 (2iK)^\ell &= A_1 \frac{\Gamma(1+2iP)\Gamma(2\ell+1)}{\Gamma(\ell+1)\Gamma(\ell+1+2iP)} (2\kappa)^{-\ell},\\
B_2 (2iK)^{-\ell-1} &= A_1 \frac{\Gamma(1+2iP)\Gamma(-2\ell-1)}{\Gamma(-\ell)\Gamma(2iP-\ell)} (2\kappa)^{\ell+1},\\
\frac{B_2}{B_1} &=  \, \frac{\Gamma(\ell+1)\Gamma(-2\ell-1)\Gamma(\ell+1+2iP)}{\Gamma(2\ell+1)\Gamma(-\ell)\Gamma(2iP-\ell)}(4i\kappa K)^{2\ell+1}. \label{B2B1}
\end{align}
These relations establish the explicit connection between the near-horizon ingoing solution and the far-region scattering solution.

With the asymptotic form of the radial solutions established, we now introduce the superradiant amplification factor $(Z)$. In the scattering setup, the scalar field at spatial infinity consists of a superposition of an incident wave propagating toward the black hole and a reflected wave propagating outward. These components can be directly identified from the asymptotic behavior \eqref{infinity}
\begin{equation}
\lim_{r\to\infty}R_\infty(r)\approx \mathcal{I}\,e^{-iKr}r^{i\Lambda-1}+\mathcal{R}\,e^{iKr}r^{-i\Lambda-1}, \label{scwave}
\end{equation}
where $\mathcal{I}$ and $\mathcal{R}$ denote the amplitudes of the incident and reflected waves, respectively,
\begin{align}
\mathcal{R} &=  \left[ B_1 \frac{\Gamma(2\ell+2)}{\Gamma(\ell+1-i\Lambda)} + B_2 \frac{\Gamma(-2\ell)}{\Gamma(-\ell-i\Lambda)} \right] {\left(2iK\right)^{-i\Lambda-1}},\\
\mathcal{I} &=\left[ B_1 {\left(-1\right)^{-\ell-1+i\Lambda}}\frac{\Gamma(2\ell+2)}{\Gamma(\ell+1+i\Lambda)} + B_2 {\left(-1\right)^{\ell+i\Lambda}}\frac{\Gamma(-2\ell)}{\Gamma(-\ell+i\Lambda)} \right]{\left(2iK\right)^{-1+i\Lambda}}.
\end{align}
This decomposition naturally leads to the definition of the amplification factor as the relative enhancement of the reflected flux with respect to the incident one,
\begin{equation}
Z=\frac{|\mathcal{R}|^2}{|\mathcal{I}|^2}-1,
\end{equation}
where a positive value of $Z$ signals amplification of the reflected wave and thus the occurrence of superradiance, whereas $Z<0$ corresponds to absorption by the black hole.

Explicitly, the ratio between reflected and incident amplitudes can be written explicitly as
\begin{equation}
\frac{|\mathcal{R}|^2}{|\mathcal{I}|^2}=\left| \frac{\left[\frac{\Gamma(2\ell+2)}{\Gamma(\ell+1-i\Lambda)}+\frac{\Gamma(-2\ell)}{\Gamma(-\ell-i\Lambda)}\left(\frac{B_2}{B_1}\right)\right]{\left(2iK\right)^{-i\Lambda-1}}} { \left[{\left(-1\right)^{-\ell-1+i\Lambda}}\frac{\Gamma(2\ell+2)}{\Gamma(\ell+1+i\Lambda)}+ {\left(-1\right)^{\ell+i\Lambda}}\frac{\Gamma(-2\ell)}{\Gamma(-\ell+i\Lambda)}\left(\frac{B_2}{B_1}\right)\right]{\left(2iK\right)^{-1+i\Lambda}}}\right|^2,
\end{equation}
where $B_2/B_1$ is given in \eqref{B2B1}. 

Employing the reflection formula for gamma function with negative argument \cite{bell1968special},
\begin{gather}
\Gamma(z)\Gamma(1-z)=\frac{\pi}{\sin (\pi z)} \quad \longrightarrow \quad    \frac{\Gamma(-a)}{\Gamma(-b)}=\frac{\sin (\pi b) \Gamma(1+b)}{\sin (\pi a) \Gamma(1+a)},
\end{gather}
and recalling that in the regime required for the validity of the AAM, one has $Mm\ll 1$, so that the parameter $\Lambda=\frac{A m^2}{2K}\sim Mm \ll 1$ remains perturbatively small, $\Lambda \ll 1$, we can approximate
\begin{gather}
    \frac{\Gamma(-2\ell)}{\Gamma(-\ell+i\Lambda)}\approx\frac{\Gamma(-2\ell)}{\Gamma(-\ell)}=(-1)^\ell \frac{\ell!}{2(2\ell)!},\\
    \frac{\Gamma(-2\ell-1)}{\Gamma(-\ell)}= (-1)^{\ell+1}\frac{\ell!}{(2\ell+1)!}. \label{approxfor}
\end{gather}


\begin{figure}[h]
    \centering
    \includegraphics[scale=0.35]{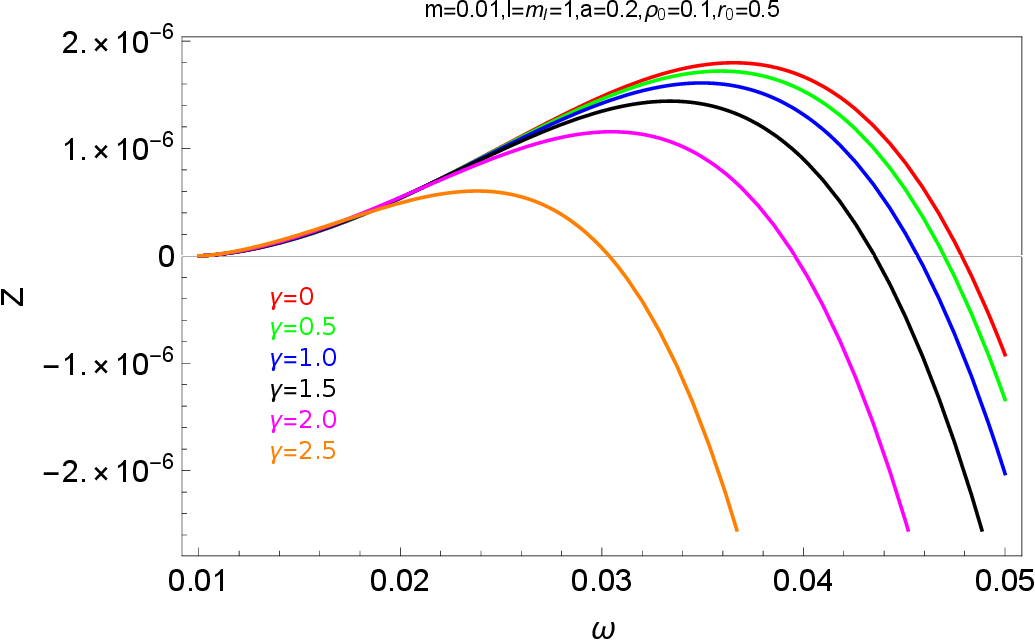}
    \includegraphics[scale=0.35]{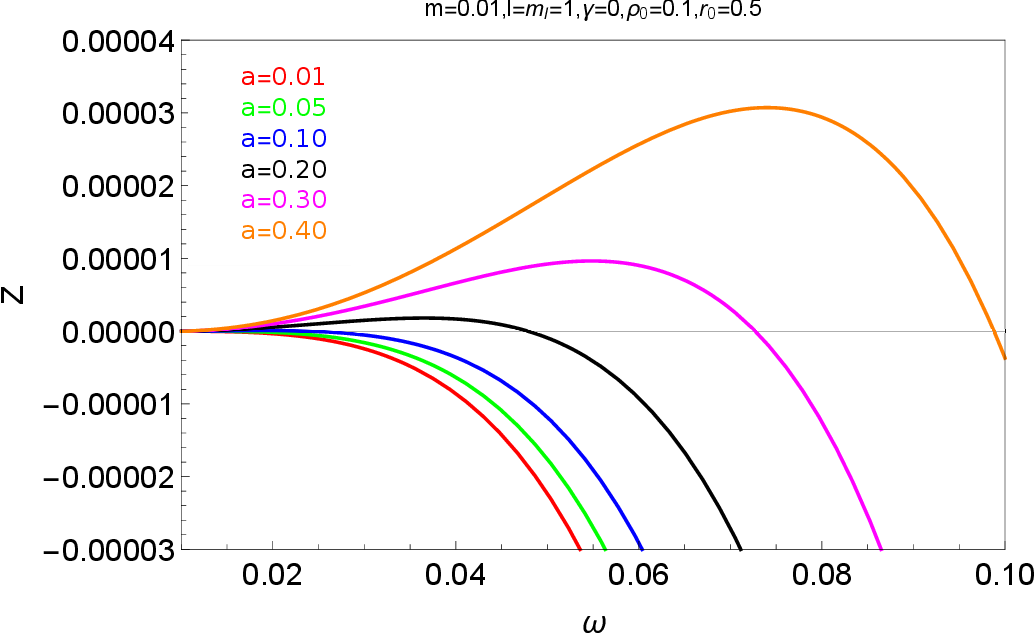}\\
    \includegraphics[scale=0.35]{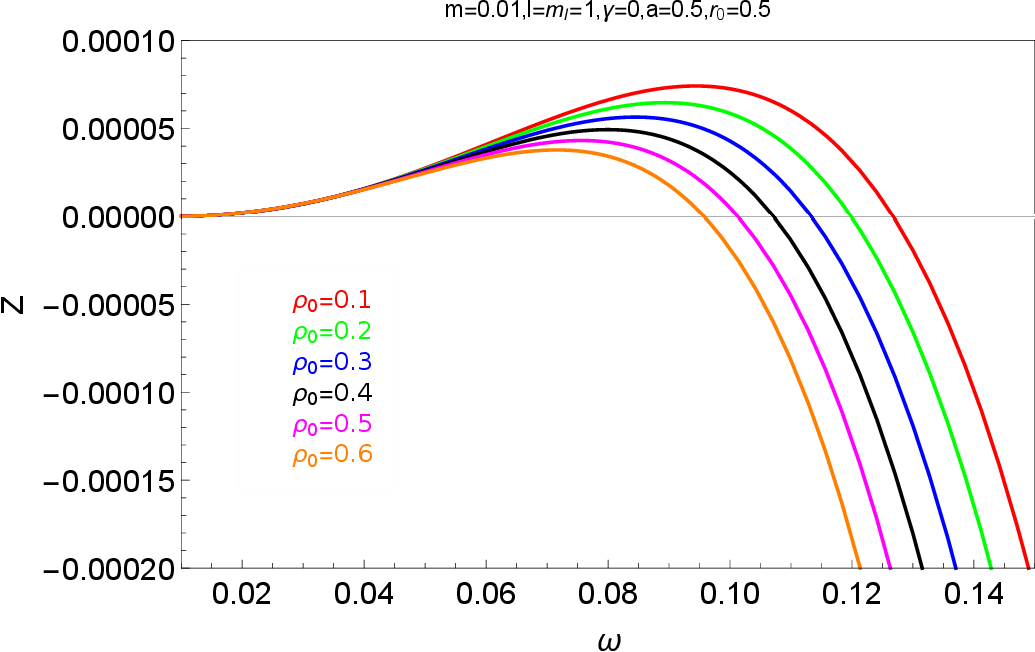}
    \includegraphics[scale=0.35]{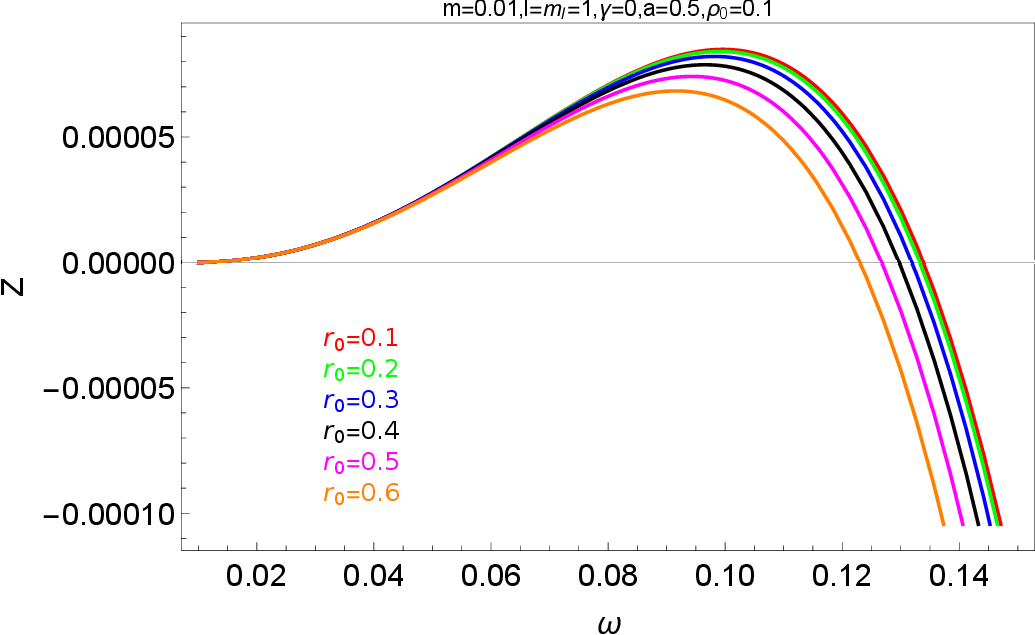}
    \caption{The amplification factor $Z$ plotted as a function of frequencies $\omega$. Each panel illustrates the effect of varying one of the four parameters, $\gamma,a,\rho_0$ and $r_0$, with the remaining parameters kept constant.} \label{fig:Z}
\end{figure}

Figure~\ref{fig:Z} illustrates the superradiant amplification factor $Z$ as a function of the scalar-field frequency for the fundamental co-rotating mode $(\ell=m_\ell=1)$. Superradiance occurs only within the frequency interval $0<\omega<m_\ell\Omega_H$, where the amplification factor is positive. In all panels, $Z$ vanishes at the critical frequency $\omega=m_\ell\Omega_H$, beyond which the scalar wave is absorbed by the black hole rather than amplified.

The upper-left panel shows the effect of varying the Dehnen profile parameter $\gamma$. As $\gamma$ increases, both the peak amplification factor and the critical frequency decrease, leading to a progressively narrower superradiant frequency window. This indicates that steeper dark matter density profiles suppress the efficiency of rotational energy extraction. A similar behavior is observed in the lower-left panel, where increasing the central dark matter density $\rho_0$ also reduces both the maximum amplification and the critical frequency. These results suggest that denser dark matter halos weaken the superradiant scattering process.

The upper-right panel illustrates the dependence on the rotation parameter $a$. As expected, the superradiant amplification becomes stronger with increasing black hole spin. Both the peak amplification and the critical frequency increase monotonically with $a$, reflecting the larger angular velocity of the event horizon and the consequently wider superradiant frequency interval.

Finally, the lower-right panel shows the influence of the characteristic halo radius $r_0$. Increasing $r_0$ shifts the critical frequency to lower values while simultaneously reducing the peak amplification factor. Consequently, larger dark matter halos suppress the efficiency of superradiant scattering in a manner qualitatively similar to increasing either $\gamma$ or $\rho_0$. Overall, the results demonstrate that more extended, denser, and cuspier dark matter halos systematically reduce both the strength and the frequency range of superradiant amplification.

\section{Thermal Superradiant Energy Extraction} \label{sec6}
In this section we consider the energy extracted from the rotating black hole through superradiant scattering. To this end, we first derive the asymptotic energy flux extracted by the scattered scalar field. We begin with the Lagrangian density of a generally complex scalar field propagating in a curved spacetime \cite{Liu_2024},
\begin{align}
\mathcal{L}=g^{\mu\nu}\partial_{(\mu}\psi^*\partial_{\nu)}\psi-\frac{1}{2}m^2|\psi|^2,
\end{align}
where
\begin{align}
\partial_{(\mu}\psi^*\partial_{\nu)}\psi=\frac{1}{2}\left(\partial_\mu\psi^*\partial_\nu\psi+\partial_\nu\psi^*\partial_\mu\psi\right).
\end{align}
The corresponding energy-momentum tensor is
\begin{align}
T^\mu_{\ \nu} = g^{\mu\lambda}\partial_{(\lambda}\psi^*\partial_{\nu)}\psi - \delta^\mu_{\ \nu}\mathcal{L}.
\end{align}
The energy flux per unit solid angle measured by an observer at spatial infinity is defined as \cite{Brito:2015oca}
\begin{align}
\frac{d^2E}{dt\,d\Omega} = \lim_{r\to\infty} r^2T^r_{\ t}.
\end{align}
Using the the asymptotic behavior of the radial function \eqref{scwave}, we re-normalize the incident wave as follows,
\begin{equation}
\lim_{r\to\infty}R_\infty(r)\approx\frac1r\left[  e^{-iKr}+\frac{\mathcal{R}}{\mathcal{I}}\,e^{iKr}\right],
\end{equation}
Recall that $\psi = e^{-i\omega t + i m_\ell\phi}R(r)S_\ell^{m_\ell}(\theta)$, therefore the mixed component of the energy-momentum tensor becomes
\begin{align}
T^r_{\ t} &= \frac{1}{2}g^{rr} \left( \partial_r\psi^*\partial_t\psi + \partial_t\psi^*\partial_r\psi \right), \nonumber \\
&= -\frac{i\omega}{2} g^{rr} \left( R_\infty R_\infty'^* - R_\infty^*R_\infty' \right) |S_\ell^{m_\ell}|^2,
\nonumber\\ 
&= \omega g^{rr} \,\mathrm{Im} \left( R_\infty R_\infty'^* \right) |S_\ell^{m_\ell}|^2. 
\end{align}
Substituting the asymptotic radial solution and retaining the leading-order contribution in $1/r$ gives \begin{align} 
T^r_{\ t}(r\rightarrow\infty) \approx \frac{\omega K}{r^2}\left|\frac{\mathcal{R}}{\mathcal{I}}\right|^2 |S_\ell^{m_\ell}|^2. 
\end{align} 
where we have used the asymptotic flatness of the spacetime, $g^{rr}\approx 1$ as $r\rightarrow\infty$.

Multiplying by $r^2$ and integrating over the solid angle using the normalization of the spheroidal harmonics, 
\begin{align}
\int |S_l^{m_l}|^2 \,d\Omega = 1, 
\end{align} 
we arrive at the outgoing energy flux 
\begin{align} \dot{E}_{\rm out} = \omega K \left|\frac{\mathcal{R}}{\mathcal{I}}\right|^2 .
\end{align}
For massless scalar, $m=0$, the asymptotic dispersion relation reduces to
\begin{align}
K=\omega.
\end{align}
In a realistic scattering process, the incident scalar field is described by a wave packet rather than by a monochromatic mode. The total extracted energy is therefore obtained by integrating over the spectral distribution of the incident field, 
\begin{align}
\dot{E}_{\rm tot} = \int_0^\infty \omega^2 \left(\left|\frac{\mathcal{R}(\omega)}{\mathcal{I}(\omega)}\right|^2 -1\right)n(\omega)\, d\omega = \int_0^\infty \omega^2 Z(\omega) n(\omega)\, d\omega,
\end{align}
where $n(\omega)$ denotes the normalized spectral weight of the incoming radiation.

Assuming that the incident field is in thermal equilibrium at temperature $T$, the occupation number per mode follows the Bose--Einstein distribution,
\begin{align}
\tilde n(\omega)=\frac{1}{e^{\omega/T}-1}.
\end{align}
To construct a normalized spectral weight, the density of states must be included. For a massless scalar field in three spatial dimensions, the number of modes in the interval $(K,K+dK)$ scales as $k^2dk$. Since $K=\omega$, the corresponding phase-space measure is $\omega^2d\omega$. The normalized spectral weight is therefore
\begin{align}
n(\omega) = \frac{\tilde n(\omega)\omega^2} {\displaystyle \int_0^\infty \tilde n(\omega)\omega^2\,d\omega}.
\end{align}
The normalization integral evaluates to
\begin{align}
\int_0^\infty \frac{\omega^2} {e^{\omega/T}-1} \,d\omega = 2\zeta(3)T^3,
\end{align}
where $\zeta(z)$ is the ordinary zeta function \cite{NIST},
\begin{equation}
    \zeta(z)=\sum_{n=1}^\infty n^{-z} =\frac{1}{\Gamma(z)}\int_0^\infty \frac{x^{z-1}}{e^x-1}dx.
\end{equation}
The normalized spectral weight now reads
\begin{align}
n(\omega)=\frac{1}{2\zeta(3)T^3}\left(\frac{\omega^2}{e^{\omega/T}-1}\right),
\end{align}
which satisfies the following normalization condition
\begin{align}
\int_0^\infty n(\omega)\,d\omega=1.
\end{align}
We note that the reflection relative probability $|\mathcal{R}(\omega)/\mathcal{I}(\omega)|^2$ is entirely determined by the black hole geometry and is independent of the temperature $T$, which only governs the occupation of the incident modes. To isolate the energy extracted through stimulated superradiant scattering from the intrinsic Hawking emission, we consider the regime $T\gg T_H$, where the incoming flux dominates over the spontaneous Hawking flux \cite{Wondrak_2018}.

\begin{figure}[h]
    \centering
    \includegraphics[scale=0.36]{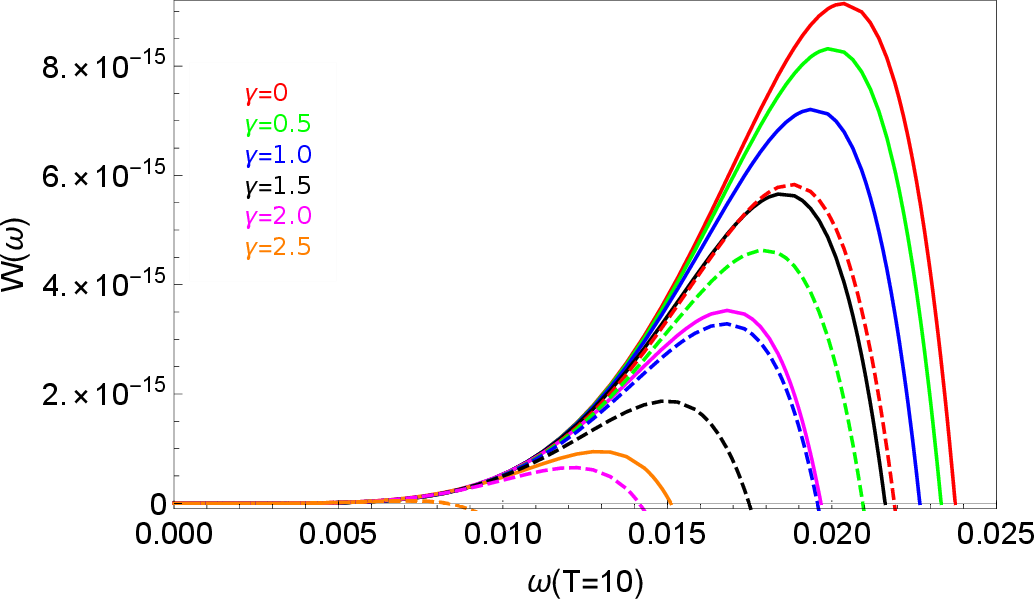}
    \includegraphics[scale=0.36]{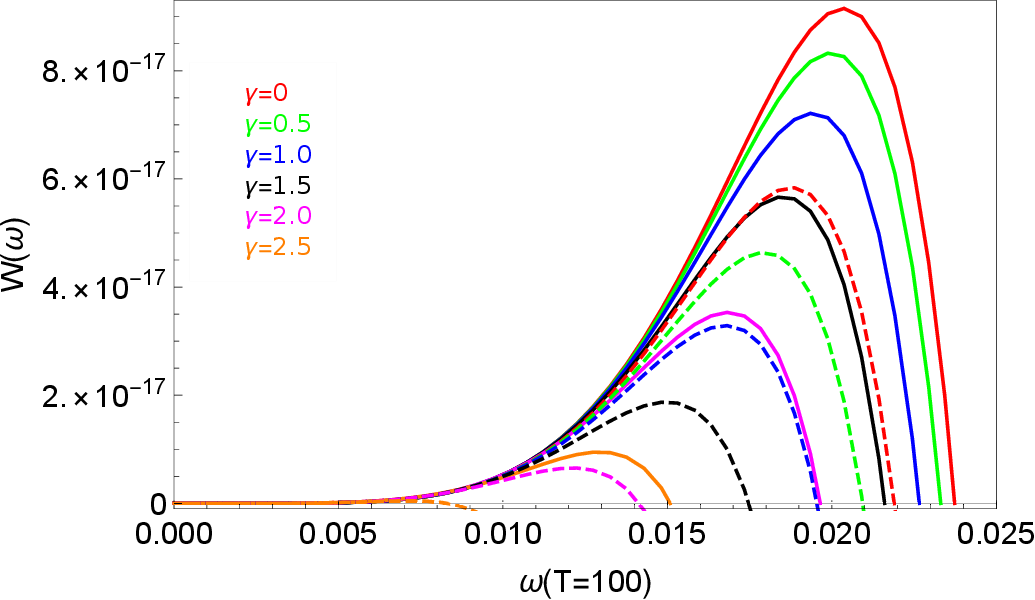}
    \caption{Energy extraction rate for for black hole with $r_s=2, r_0=0.5, a=0.1, \ell=m_\ell=1$ and $T=10, 100$. The solid/dashed lines denote $\rho_0=0.1$ and $\rho_0=0.25$ respectively.  } \label{Spec1}
\end{figure}

The spectral energy extraction rate is therefore
\begin{align}
W(\omega) \equiv 2\zeta(3) \frac{d\dot E_{\rm out}}{d\omega} = \frac{Z(\omega)} {T^3} \left( \frac{\omega^4} {e^{\omega/T}-1} \right).
\end{align}
In Fig.~\ref{Spec1}, we present the spectral energy extraction rate $W(\omega)$ for different values of the Dehnen profile parameter $\gamma$. Throughout, we fix $r_s=2$, $r_0=0.5$, $a=0.1$, and $\ell=m_\ell=1$. Two dark matter densities, $\rho_0=0.1$ and $\rho_0=0.25$, together with two thermal bath temperatures, $T=10$ and $T=100$, are considered to illustrate the influence of the halo properties and the incident radiation.

For all parameter choices, the spectrum is governed by the competition between the thermal occupation of the incident photons and the superradiant amplification factor $Z(\omega)$. Consequently, energy extraction occurs only within the superradiant window, $\omega<m_\ell\Omega_H$, and decreases continuously to zero as the frequency approaches the critical value $\omega=m_\ell\Omega_H$. The effect of the dark matter profile is evident: smaller values of $\gamma$, corresponding to less centrally concentrated (less cuspy) Dehnen halos, produce a larger spectral energy extraction rate, whereas increasing $\gamma$ progressively suppresses the spectrum.

The temperature primarily controls the distribution of the incident thermal photons. At lower temperatures, the thermal spectrum is concentrated at lower frequencies, where the overlap with the superradiant band is greatest, resulting in a larger extracted power. As the temperature increases, the thermal distribution shifts toward higher frequencies, placing a greater fraction of incident photons outside the superradiant regime and thereby reducing the overall energy extraction rate.

A similar suppression is observed when the halo density is increased from $\rho_0=0.1$ to $\rho_0=0.25$. For fixed $\gamma$, a denser dark matter halo consistently lowers the spectral energy extraction rate over the entire superradiant frequency range. This behavior reflects the stronger gravitational influence of the surrounding halo, which weakens the efficiency of rotational energy extraction from the black hole.

\section{Conclusion and Discussion} \label{sec7}
We have successfully constructed a general Kerr-like black hole embedded in a double power law dark matter halo by applying the Newman--Janis algorithm to a Schwarzschild-like seed metric. The resulting spacetime provides a unified rotating solution applicable to an arbitrary double power law density profile, thereby extending several previously known rotating black holes associated with specific dark matter models. This construction offers a general framework for investigating the interplay between black hole rotation and realistic galactic dark matter environments.

The geometry exhibits several interesting features. We found that the presence of the dark matter halo modifies the locations of the event horizons, the extremal spin, and the stationary-limit surfaces. In particular, steeper inner density profiles allow the black hole to attain higher angular momentum before reaching extremality. A remarkable result is that the rotating spacetime becomes free of essential curvature singularities for Dehnen profiles with $\gamma\le2$, despite the corresponding static seed solution remaining singular. This demonstrates that the combined effects of rotation and the surrounding dark matter distribution can significantly alter the local structure of the spacetime.

Using this geometry as the background, we investigated massive scalar perturbations by specializing to the Dehnen $(1,4,\gamma)$ dark matter profile. Employing the analytical asymptotic matching method, we derived analytical expressions for both the quasibound spectrum and the superradiant amplification factor in the low-frequency regime. We found that the dark matter halo leaves characteristic signatures on the scalar spectrum through the combined parameter $\rho_0r_0^3$ and the inner density slope $\gamma$. In the quasibound state regime, increasing either $\rho_0r_0^3$ or $\gamma$ produces more tightly bound states with larger binding energies while simultaneously increasing the magnitude of the imaginary part of the frequency, corresponding to shorter-lived quasibound states. Furthermore, the threshold condition for scalar cloud formation is modified through the horizon quantity $\xi(r_H)$, leading to a shift of the critical cloud mass relative to the Kerr case. 

In the scattering regime, increasing either $\rho_0r_0^3$ or $\gamma$ suppresses the superradiant amplification by lowering both the critical frequency $\omega=m_\ell\Omega_H$, which forms the upper boundary of the superradiant interval $0<\omega<m_\ell\Omega_H$, and the maximum amplification factor. Consequently, the superradiant frequency window becomes progressively narrower. This behavior occurs exclusively for co-rotating modes ($m_\ell>0$), whereas counter-rotating modes ($m_\ell<0$) do not satisfy the superradiant condition and therefore exhibit no amplification.

Furthermore, we investigated the extraction of rotational energy driven by thermal scalar fields. Starting from the scalar energy-momentum tensor, we derived the asymptotic energy flux and obtained an analytical expression for the thermally averaged spectral energy extraction rate. We found that the extracted energy is restricted to the superradiant frequency band and is determined by the combined effects of the Bose--Einstein distribution and the superradiant amplification factor. Our numerical results show that smaller values of the Dehnen profile parameter $\gamma$ produce a higher energy extraction rate, while increasing the central dark matter density $\rho_0$ suppresses the overall extracted power. Furthermore, lower temperatures shift the dominant contribution of the thermal spectrum toward lower frequencies, increasing its overlap with the superradiant regime and enhancing the extracted power.

\section*{Acknowledgment}
SP acknowledge funding support from the NSRF via the Program Management Unit for Human Resource and Institutional Development, Research and Innovation grant number $B39G690007$.


\appendix
\section{The Gauss Hypergeometric Equation}\label{AppendixA}

The Gauss hypergeometric equation is the prototypical second–order equation with three regular singular points~\cite{Bell},
\begin{equation}
x(1-x)\frac{d^2\psi_G}{dx^2}+\left[a_3-(a_1+a_2+1)x\right]\frac{d\psi_G}{dx}-a_1a_2\psi_G=0.
\label{GaussCanonical}
\end{equation}
Its general solution is
\begin{equation}
\psi_G=A\,{}_2F_1(a_1,a_2,a_3,x)+B\,x^{1-a_3}{}_2F_1(a_1-a_3+1,a_2-a_3+1,2-a_3,x),
\end{equation}
where $A$ and $B$ are constants. Following Appendix~\ref{AppendixC}, equation \eqref{GaussCanonical} can be cast into normal form by defining
\begin{equation}
\Psi_G=x^{\frac{a_3}{2}}(1-x)^{\frac{1}{2}(1+a_1+a_2-a_3)}\psi_G,
\end{equation}
which yields
\begin{equation}
\frac{d^2\Psi_G}{dx^2}-V_{\rm eff}(x)\Psi_G=0,
\label{GaussNormal}
\end{equation}
with
\begin{equation}
V_{\rm eff}(x)=\frac{x^2\!\left((a_1-a_2)^2-1\right)-2x\!\left((a_1+a_2-1)a_3-2a_1a_2\right)+a_3(a_3-2)}{4x^2(1-x)^2}.
\end{equation}

A key ingredient in asymptotic matching is the analytic continuation of ${}_2F_1$ to large argument. The connection formula is given by
\begin{multline}
{}_2F_1(a_1,a_2,a_3,x)=
\frac{\Gamma(a_2-a_1)\Gamma(a_3)}{\Gamma(a_2)\Gamma(a_3-a_1)}(-x)^{-a_1}
{}_2F_1\!\left(a_1,a_1-a_3+1,a_1-a_2+1,\frac{1}{x}\right)\\
+\frac{\Gamma(a_1-a_2)\Gamma(a_3)}{\Gamma(a_1)\Gamma(a_3-a_2)}(-x)^{-a_2}
{}_2F_1\!\left(a_2,a_2-a_3+1,a_2-a_1+1,\frac{1}{x}\right),
\label{GaussConnection}
\end{multline}
which is the central tool used to extract the large-$x$ behaviour of the near-horizon solution in the main text.

\section{The Confluent Hypergeometric Equation}\label{AppendixB}

The confluent hypergeometric equation arises as a canonical normal form in a wide class of radial problems. It can be written as~\cite{Bell}
\begin{equation}
\frac{d^2\psi_C}{dx^2}+\left(-\frac{1}{4}+\frac{k}{x}+\frac{\frac{1}{4}-m^2}{x^2}\right)\psi_C=0,
\label{WhittakerNormal}
\end{equation}
which is recognized as the Whittaker equation.

Its general solution is given in terms of the Whittaker functions $M_{k,m}$
\begin{equation}
\psi_C=A\,M_{k,m}(x)+B\,M_{k,-m}(x),
\end{equation}
where $A$ and $B$ are constants. The Whittaker functions are related to the confluent hypergeometric function ${}_1F_1$ via
\begin{equation}
M_{k,\pm m}(x) = x^{\frac{1}{2}\pm m} e^{-\frac{x}{2}} 
{}_1F_1 \left( \frac{1}{2}-k\pm m, 1\pm 2m, x \right).
\end{equation}

The power-law prefactor controls the behaviour near the origin, while the exponential factor governs the asymptotics at large $x$. For $|x|\to\infty$, one has~\cite{NIST}
\begin{multline}
{}_1F_1(a,b,x)=\frac{\Gamma(b)}{\Gamma(a)}e^{x}x^{a-b}
{}_2F_0\left(b-a,1-a;-\frac{1}{x}\right)\\
+\frac{\Gamma(b)}{\Gamma(b-a)}(-x)^{-a}
{}_2F_0\left(a,a-b+1;-\frac{1}{x}\right).
\label{HyperAsymptotic}
\end{multline}

For integer parameters $a,b\in\mathbb{N}$, let $\epsilon$ be a small complex parameter satisfying $|\epsilon|\ll1$. Expanding the confluent hypergeometric function about the pole $-a$ to first order in $\epsilon$, one obtains the small-$x$ asymptotic form \cite{Abramowitz1964,Detweiler:1980uk,Furuhashi,Yang22}
\begin{equation}
{}_1F_1(-a-\epsilon,b,x)
\approx
(-1)^a\frac{\Gamma(a+b)}{\Gamma(b)}
+\epsilon\,(-1)^{a+1}\Gamma(a+1)\Gamma(b-1)\,x^{\,1-b},
\label{1F1_small_x_eps}
\end{equation}
The second term arises from the first-order Taylor expansion in $\epsilon$ around the integer value $-a$.

\section{Normal Form}\label{AppendixC}
A second–order ordinary differential equation can be cast into a normal (Schrödinger-like) form by eliminating the first–derivative term. This representation is particularly useful for qualitative and asymptotic analysis~\cite{NIST}.

Now, we consider the linear equation
\begin{equation}
\frac{d^2 y}{dx^2}+p(x)\frac{dy}{dx}+q(x)y=0.
\label{generalODE}
\end{equation}
To remove the first–derivative term, we introduce the transformation
\begin{equation}
y(x)=Y(x)\exp\!\left(-\frac{1}{2}\int p(x)\,dx\right),
\end{equation}
which is constructed such that the resulting equation for $Y(x)$ contains no $dY/dx$ term~\cite{2420}. Substituting into \eqref{generalODE}, we obtain
\begin{equation}
\frac{d^2 Y}{dx^2} +\left( -\frac{1}{2}\frac{dp}{dx} -\frac{1}{4}p^2 +q \right)Y=0.
\label{normalform}
\end{equation}
By defining,
\begin{equation}
Q(x)=-\frac{1}{2}\frac{dp}{dx}-\frac{1}{4}p^2+q,
\end{equation}
equation \eqref{normalform} takes the compact form
\begin{equation}
\frac{d^2 Y}{dx^2}=-Q(x)Y.
\end{equation}

This representation provides immediate qualitative insight. If $Q(x)>0$, the solution is locally oscillatory and necessarily crosses the $x$-axis. Moreover, if
\begin{equation}
\int^{\infty} Q(x)\,dx = \infty,
\end{equation}
then $Y(x)$ possesses infinitely many zeros~\cite{2420}. In contrast, if $Q(x)<0$, the solution is non-oscillatory and admits at most one zero.

\bibliography{sn-bibliography}


\end{document}